\documentclass{aa}

\usepackage{rotating}
\usepackage{graphicx}
\usepackage{sidecap}
\usepackage{subcaption}
\usepackage{url}
\usepackage[colorlinks=true]{hyperref}
\hypersetup{allcolors=blue}
\usepackage{microtype} 
\usepackage{lipsum}
\usepackage{txfonts}
\usepackage{float}
\usepackage{natbib}
\bibpunct{(}{)}{;}{a}{}{,}
\DeclareUnicodeCharacter{00A0}{~}
\usepackage[us,12hr]{datetime}
\usepackage{ulem}

\newcommand{\vect}[1]{\boldsymbol{\mathbf{#1}}}

\usepackage{cancel}
\usepackage{color}
\usepackage{tikz}

\definecolor{deepmagenta}{rgb}{0.8, 0.0, 0.8}
\definecolor{green}{rgb}{0.0, 0.5, 0.1}

\patchcmd{\thebibliography}{\clubpenalty4000}{\clubpenalty10000}{}{}
\patchcmd{\thebibliography}{\widowpenalty4000}{\clubpenalty10000}{}{}

\begin{document}

    \title{Approximate Bayesian Neural Doppler Imaging}

   \author{A. Asensio Ramos
          \inst{1,2}
          \and
          C. J. D\'iaz Baso
          \inst{3}
          \and
          O. Kochukhov
          \inst{4}
          }

   \institute{Instituto de Astrof\'isica de Canarias, C/V\'{\i}a L\'actea s/n, E-38205 La Laguna, Tenerife, Spain \email{andres.asensio@iac.es}
   \and Departamento de Astrof\'isica, Universidad de La Laguna, E-38206 La Laguna, Tenerife, Spain
   \and Institute for Solar Physics, Dept. of Astronomy, Stockholm University, AlbaNova University Centre, SE-10691 Stockholm, Sweden
   \and Department of Physics and Astronomy, Uppsala University, Box 516, Uppsala 75120, Sweden}

   \date{Draft: compiled on \today\ at \currenttime~UT}

   \authorrunning{Asensio Ramos et al.}


   \abstract
   {}
   {The non-uniform surface temperature distribution of rotating active stars is routinely mapped with the 
   Doppler Imaging technique. Inhomogeneities in the surface produce features in high-resolution spectroscopic
   observations that shift in wavelength due to the Doppler effect depending on their position on the
   visible hemisphere. The inversion problem has been systematically solved using
   maximum a-posteriori regularized methods assuming smoothness or maximum entropy. Our aim in this work
   is to solve the full Bayesian inference problem, by providing access to the posterior distribution
   of the surface temperature in the star compatible with the observations.}
   {We use amortized neural posterior estimation to produce a model that approximates the high-dimensional posterior
   distribution for spectroscopic observations of selected spectral ranges sampled at arbitrary rotation
   phases. The posterior distribution is approximated with conditional
   normalizing flows, which are flexible, tractable and easy to sample approximations to
   arbitrary distributions. When conditioned on the spectroscopic observations, they provide a very 
   efficient way of obtaining samples from the posterior distribution. The conditioning on observations
   is obtained through the use of Transformer encoders, which can deal with arbitrary wavelength 
   sampling and rotation phases.}
   {Our model can produce thousands of posterior samples per second, each one accompanied with an estimation
   of the log-probability. Our exhaustive validation of the model for very high signal-to-noise
   observations shows that it correctly approximates the posterior, although with some overestimation of the broadening.
   We apply the model to the moderately fast rotator II Peg, producing the first Bayesian map of its
   temperature inhomogenities. We conclude that conditional normalizing flows are a very promising tool to carry out
   approximate Bayesian inference in more complex problems in stellar physics, like 
   constraining the magnetic properties using polarimetry.}
   {}


   \maketitle


\section{Introduction}\label{sec:intro}
Different classes of stars with non-uniform surface structures exhibit rotational modulation 
of their spectra. This variability is produced by the changing visibility of surface spots in 
the course of stellar rotation. When observed at high spectroscopic resolution, Doppler 
broadened line profiles of these stars show characteristic distortions -- bumps or dips -- propagating 
across the line as the star rotates. This Doppler resolution of stellar surfaces enables 
one to recover a wealth of information about the topology and evolution of surface spots 
with the technique known as Doppler Imaging \citep[DI, e.g.][]{rice2002,kochukhov2016}.

The first practical application of the spectroscopic starspot mapping dates back to the 
work of \citet{deutsch58}, who used a harmonic expansion method to study chemical abundance 
distributions on early-type chemically peculiar stars. Other investigations attempted to 
map spots on these stars with a trial and error approach \citep{khokhlova1976}. In a key 
development, \citet{goncharskii1977,goncharsky82} proposed a method to recover unparametrised 
two-dimensional chemical maps of early-type stars by recasting DI as a regularised ill-posed 
inversion problem. Subsequently, this version of DI was extended to mapping brightness 
\citep{collier-cameron1986,vogt87} and temperature \citep{piskunov1990,berdyugina1998,rice2000} 
inhomogeneities on the surfaces of active late-type stars and was applied to the reconstruction 
of isotopic composition maps \citep{adelman2002}, non-radial pulsational velocity field 
\citep{kochukhov2004}, and even clouds in brown dwarf atmospheres \citep{crossfield2014}.

With the advent of high-resolution spectropolarimetry it became possible to apply the 
principles of tomographic mapping not only to intensity but also to polarization spectra 
in order to reconstruct vector maps of stellar surface magnetic fields \citep{piskunov1983,semel89,brown91,piskunov2002}. 
This technique, known as Zeeman (Magnetic) Doppler Imaging (ZDI), is actively used today to 
study magnetic fields of essentially all classes of stars \citep[e.g.][]{folsom2018,kochukhov2019,strassmeier2019} 
and currently represents the only viable approach for obtaining information about magnetic field 
topologies of stars other than the Sun.

Similar to almost any inversion problem in physics, DI and ZDI inversions are ill-posed. 
This means that information in observational data is often insufficient for unique parameter 
inference. To counteract this issue, additional constraints, often based on consideration of 
simplicity, smoothness or least amount of information, have to be implemented to ensure 
convergence of spectral fitting to a unique solution and stability of this solution against 
random noise in the data. The two most popular regularisation strategies employed in DI is 
the Tikhonov regularisation \citep{tikhonov1977} and the maximum entropy method \citep{skilling84}. 
On the other hand, most modern implementations of ZDI rely on a spherical harmonic decomposition 
of stellar magnetic field and implement regularisation by truncating this harmonic expansion 
or penalising its higher-order harmonic terms \citep[e.g.][]{hussain2000,kochukhov2014}.

Despite the enormous success of regularized optimization methods, we still
lack a fully probabilistic solution to the tomographic mapping problem. We propose
in this work a Bayesian solution to DI that is also extremely efficient thanks to
the use of recent tools developed in the field of deep learning. Our focus in this paper is on 
DI but this can be easily extended to ZDI and abundance mapping (which might
even include physically realistic constraints).

\section{Amortized neural posterior estimation}
\label{sec:anpe}
Given that the DI problem is ill-defined, one needs to resort to a probabilistic
description and solve the inference problem in the Bayesian framework \citep[see, e.g.,][]{gregory05}. 
Assuming an arbitrary pixelation of the
surface of the star, our aim is to obtain the vector $\mathbf{T}$ of all temperatures
in the surface of the star that are compatible with the observations. In general, 
the observations consist of the spectrum of the star, either individual spectral lines, 
a set of spectral regions or even 
an average line profile
for cases with reduced signal-to-noise ratio (SNR), for different
rotational phases of the star. We use the generic notation $I_i(\lambda)$ to refer to 
the observed spectrum at the $i$-th rotation phase. For notational convenience, we use 
$D=\{I_i(\lambda), i=1\ldots N\}$ to represent the spectra obtained at all $N$ observed 
rotation phases of the star. The fully Bayesian solution of the DI problem requires the computation of $p(\mathbf{T}|D)$, the
posterior distribution for $\mathbf{T}$ conditioned on $D$. The posterior distribution can
be calculated by a direct application of the Bayes theorem \citep{bayes}:
\begin{equation}
    p(\mathbf{T}|D) = \frac{p(D|\mathbf{T}) p(\mathbf{T})}{p(D)},
\end{equation}
where $p(D|\mathbf{T})$ is the likelihood, that describes the generation of the
observations conditioned on the temperature on the surface of the star, while $p(\mathbf{T})$
is the prior distribution of all possible surface temperatures. The quantity $p(D)$ is the
so-called evidence or marginal posterior, that transforms the posterior a properly normalized
probability distribution. The evidence is non-important in parameter estimation with
Bayesian inference so we drop it in the following.

All information available for $\mathbf{T}$ is encoded on the posterior distribution.
From this posterior, one can extract summaries like the most probable surface temperature,
uncertainties and correlations. However, when the dimensionality of $\mathbf{T}$ is large because one is interested in
a very fine description of the surface of the star, the posterior distribution is 
a very high-dimensional beast. Obtaining point estimates like the maximum a-posteriori (MAP) solution
is relatively straightforward. Once a proper likelihood and priors are defined, the MAP solution
can be computed with standard optimization methods. From a practical point of view, it is customary to 
optimize instead the log-posterior (the logarithm is a monotonic operation and does not affect the 
location of the maxima):
\begin{equation}
    \log p(\mathbf{T}|D) = \log p(D|\mathbf{T}) + \log p(\mathbf{T}).
\end{equation}

The most widespread assumption to compute the MAP solution is to assume that the observations are corrupted with
uncorrelated Gaussian noise. In this case, and assuming that $S_i(\lambda_j,\mathbf{T})$ is the 
synthetic forward model of choice to fit the observations at phase $i$ and wavelength sample 
$\lambda_j$, the log-likelihood can be written as:
\begin{equation}
    \log p(D|\mathbf{T}) = -\frac{1}{2} \sum_{i=1}^N \sum_{j=1}^{N_\lambda} \left[ 
    \frac{\left( S_i(\lambda_j,\mathbf{T}) - I_i(\lambda_j) \right)^2}{\sigma_{ij}^2} - \frac{1}{2} \log 2 \pi \sigma_{ij}^2 \right],
\end{equation}
where $\sigma_{ij}^2$ is the phase- and wavelength-dependent noise variance. The number of wavelength points 
we consider is $N_\lambda$.
Typically, the forward model computes the local profiles in every point in the visible hemisphere of the
rotating star at each phase and adds them
together to produce the observed spectrum. Additional effects on the spectra produced by the
limited spectral resolution of the spectrograph are customary included via convolution with an instrumental profile.

The MAP solution is then computed by maximizing the log-posterior, with an appropriate selection
of the prior term, $\log p(\mathbf{T})$. All current DI methods (and their magnetic counterparts aimed at modeling
polarimetry) can be understood as MAP solutions
to the problem and they only differ on the specific details of the observations, forward model and
prior assumptions. It is a well-known fact that priors are needed for the solution of this 
problem because the observations do not necessarily strongly constrain the solution simply through the
likelihood. Different types of priors, specifically the maximum entropy prior or the Tikhonov 
regularization can be encoded using well-known functional forms
for $\log p(\mathbf{T})$.

Despite the success of these approaches, some challenges still remain to be solved. First, 
point estimates can be biased and might not be fully representative of the solution,
especially in complex cases. One of the pathological cases can be multimodal solutions produced
by strong ambiguities in the forward problem. 
Second, it is not always possible to obtain reliable uncertainties or correlations for DI, so it 
may be difficult to assess the reliability of the solution. It is
important to know, for instance, the uncertainty on the position and size of a predicted starspot. Finally, all MAP approaches use a simplified
likelihood under the assumption of uncorrelated noise. However, there might be additional 
latent variables that we know how to model (like small-scale activity, effects of stellar inclination, 
projected rotational velocity and differential rotation, continuum normalization etc.) but are extremely hard 
or impossible to incorporate on the likelihood function.

Approximate Bayesian Computation \citep[ABC;][]{rubin84,Beaumont02} offers an avenue to deal with many of these complications
opening the possibility to do Bayesian inference with simulations, appropriate for those cases
in which a likelihood cannot be properly defined (either intrinsically or for computational limitations).
Although classical approaches to ABC have been successful, they suffer from some problems
related to their inefficiency in large-scale problems \citep{cranmer20}. Recent advances in machine learning
(especially with the irruption of deep learning) have provided the machinery to
efficiently deal with these large-scale problems. We leverage in this paper the ideas behind amortized neural
posterior estimation \citep[ANPE;][]{cranmer20}. The aim is to train a complex neural network that directly
learns the posterior distribution $p(\mathbf{T}|D)$. We model it using 
normalizing flows \citep{flows_kobyzev20}, which we describe in the following, and use simulations for
training. These simulations can incorporate any arbitrary latent process described by generic variables $z$, so that the
posterior distribution of interest is obtained by marginalizing over the latent variables:
\begin{equation}
    p(\mathbf{T}|D) = \int p(\mathbf{T},z|D) dz = \int p(D|\mathbf{T},z) p(z|\mathbf{T}) p(\mathbf{T}) dz.
\end{equation}
As said, these latent processes can range from stochastic activity in the star in the form of
small-scale dark structures characterized by a probability distribution, to systematic effects
in the spectrograph and telescope. As long as proper simulations of these processes can be carried out, they
can be incorporated in our approach. The resulting neural network will produce inferences marginalizing
over all these uninteresting processes.

\section{Synthetic stars and their spectra}
\label{sec:stars}
We model the surface temperature of the star using the Hierarchical Equal Area and isoLatitute Pixelation
(HEALPix\footnote{\texttt{http:///healpix.sourceforge.net}}). HEALPix produces pixels of 
equal area and efficient software to deal with such grids is available. For this reason, it is the
standard for the pixelation of the sphere, especially in astrophysics \citep[e.g.,][]{2020A&A...641A...1P}.
We use $N_\mathrm{side}=16$, which results in a total of 3072 pixels distributed on the surface of the star.
This number of pixels is high enough to avoid artifacts in the synthetic spectra and low enough to
allow an affordable computation of the training set. In our case, we generate 100k random surface temperatures. 

The surface temperature maps are obtained following a recipe, which then needs to be considered 
as a prior. As such, the posterior distribution produced by our model depends on it. Anyway, we point out
that this prior can be easily updated by changing the recipe, which can be as complex as needed. Adding
complexity is especially relevant for cases like ZDI, where the additional constraint of zero-divergence
for the magnetic field has to be fulfilled. Other additional physical constraints can be imposed.

Stars are initialized by randomly choosing a temperature $T_\mathrm{star}$ uniformly between 4000 K
and 5500 K. The number of spots is randomly selected between 1 and 10. Each spot is randomly
placed on the surface of the star with no preferred latitude. The spot is chosen to be
centered on the random position and with a radius randomly extracted from a triangular distribution 
between 0.1 and 1 rad (6 to 57\degr) with a mode equal to 0.1 (this generates more small spots than
large ones). The temperature of the spot is chosen randomly between 3200 K and $1.2 T_\mathrm{star}$, limited
to a maximum of 5800 K. This generates spots with temperatures lower or higher than $T_\mathrm{star}$.

The synthetic spectra of each star at an arbitrary number of $N$ rotation phases $\{\phi_1,\ldots,\phi_N\}$
is computed by assuming that the emergent radiation at each point in the observable
hemisphere is obtained from a model atmosphere with the temperature of the pixel extracted
from the MARCS grid \citep{marcs08}. Spectra are 
computed under the local thermodynamical equilibrium assumption using a microturbulent velocity 
of 2~km\,s$^{-1}$. Every star
rotates with a rotation velocity $v_\mathrm{rot}$ randomly selected between 10 and 80 km s$^{-1}$. 
The inclination angle of
the rotation angle (with $i=0^\circ$ a star with the rotation axis pointing to the observer) is 
chosen isotropically between 10$^\circ$ and 85$^\circ$, distributed 
uniformly in $\cos i$. 
The case of $i$ and $180\degr-i$ cannot be distinguished in most DI problems (with the exception of 
the ZDI that uses four Stokes parameter spectra, e.g. \citealt{kochukhov2019}), so we choose $i<90\degr$.
The number of observed phases is also chosen randomly from a minimum of 6 and a maximum of 20.

\begin{figure}
    \centering
    \includegraphics[width=\columnwidth]{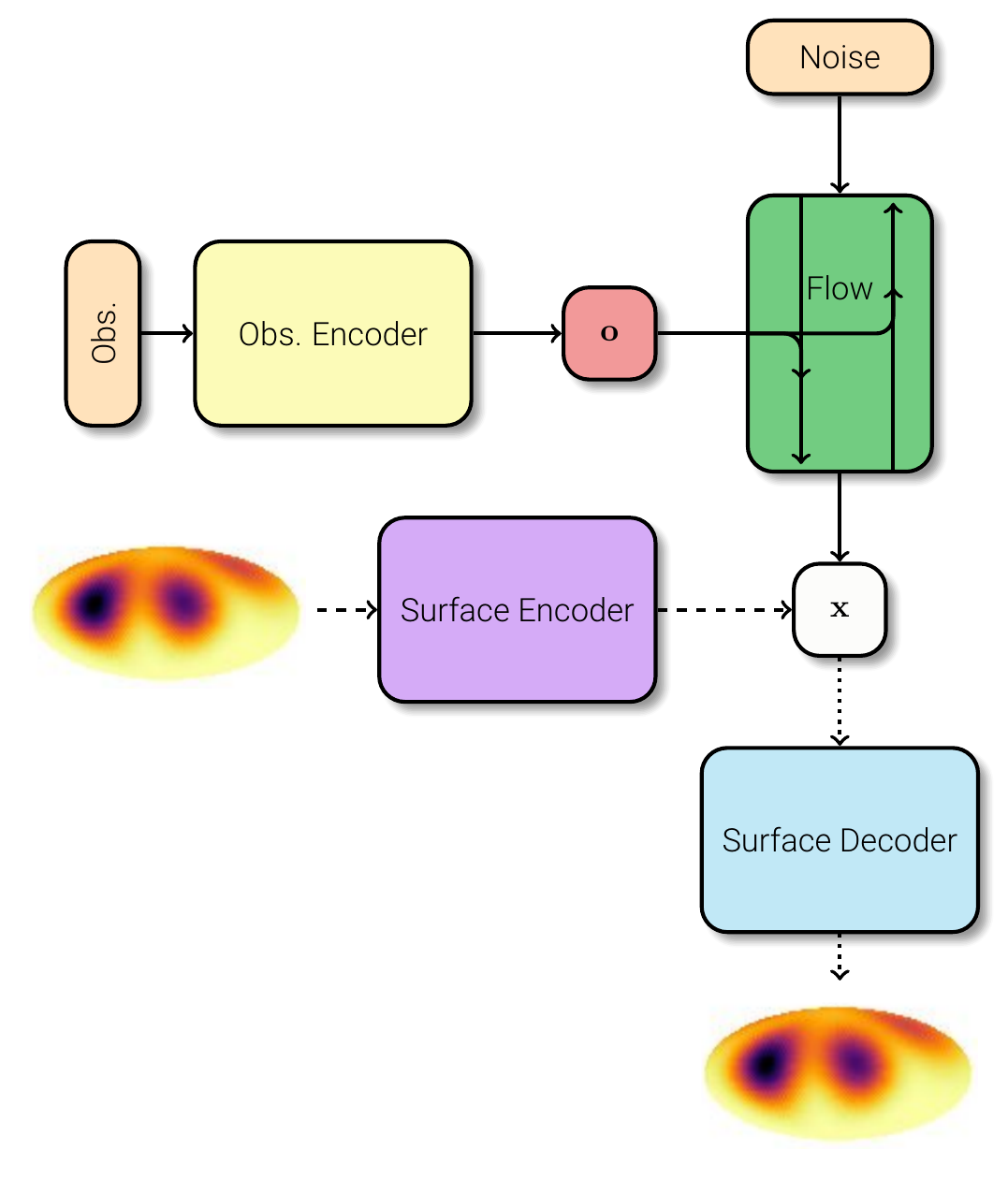}
    \caption{Schematic representation of the model. The blocks are described in detail in
    Sect. \ref{sec:model}. Solid lines show connections between blocks propagating
    gradients during training of the flow and observational encoder. 
    Dashed lines in the autoencoder do not backpropagate gradients during training. 
    Dotted lines are only used in validation to produce
    surface temperature maps.
    \label{fig:model}}
\end{figure}

In order to accelerate the computation of the spectra for the 100k stars, we precompute the MARCS
spectra for 19 temperatures between 3000 K and 6000 K (with steps of 100 K from 3000 K to 4000 K and
in steps of 250 K from 4000 K to 6000 K). Spectra for each temperature are computed at 15 astrocentric angles ($\mu$)
between 0.02 and 1 and 160 Doppler shifts in the range $[-80,80]$ km\,s$^{-1}$. This 
allows us to calculate the emergent spectrum in each pixel by a simple tri-linear interpolation, which
largely accelerates the process. The synthesis of the 100k stars with our non-optimized Python code 
takes $\sim$4 h in 40 CPU cores.

Given that we apply our model to observations from II~Peg in Sect. \ref{sec:iipeg}, we
degrade the synthetic spectra to mimic the observations with the 
ESPaDOnS spectropolarimeter \citep{donati03} at the Canada-France-Hawaii Telescope
(CFHT). We first synthesize the spectrum for the three \ion{Fe}{i} lines and relevant blends in the spectral 
regions at $(5985.1, 5989.0)$ \AA, $(6000.1, 6005.0)$ \AA, and $(6022.0, 6026.0)$ \AA, with a sampling of 0.01 \AA.
Later, they are degraded modeling an observation with a spectral resolution of $R=65,000$. This is
done by convolving the emergent spectra with a Gaussian kernel of the appropriate width. The resulting
spectra are resampled at a wavelength step of 0.03 \AA, appropriate for ESPaDOnS.
Finally, uncorrelated Gaussian with zero mean and standard deviation of $10^{-3}$ is added to the
spectrum.

\section{The model}
\label{sec:model}
The model proposed in this work for solving the DI problem is shown schematically in Fig. \ref{fig:model}.
Although we describe all the individual components of the model in detail in this section, 
let us first discuss it in general. The main component of the model is the 
normalizing flow (green block), that uses flexible neural networks to transform
Gaussian noise into samples from the posterior. We describe it in detail in Sect. \ref{sec:flow}.
This flow is conditioned on a context vector that serves as a summary of the observations. 
This encoder (yellow block) is complex because it needs to
deal with spectral lines observed at an arbitrary number of rotation
phases. This encoder is described in Sect. \ref{sec:transformer}. Given the large dimensionality 
of the stellar surface temperature maps, we
compress them using a pre-trained autoencoder, discussed in Sect. \ref{sec:autoencoder}. During the training 
of the flow and observational
encoder, surface temperature maps will be encoded with the pre-trained encoder (violet block) and
compared with outputs of the normalizing flow. During evaluation, samples from
the normalizing flow will be transformed into surface temperature maps with the
pre-trained decoder (blue block).

\subsection{Normalizing flows}
\label{sec:flow}

As described in Sect. \ref{sec:anpe}, our aim is to directly model the posterior
distribution $p(\mathbf{T}|D)$. To this end, we leverage normalizing flows
as a very flexible, tractable and easy to sample family of generative models, that can
approximate complex distributions. Simply put, a normalizing flow is a transformation of a simple
probability distribution (often a multivariate standard normal distribution, with zero
mean and unit covariance) into the desired probability distribution. Normalizing flows accomplish this by
the application of a sequence of invertible and differentiable variable transformations. Let us
assume that $\mathbf{Z}$ is a $d$-dimensional random variable with a simple and tractable probability
distribution $q_\mathbf{Z}(\mathbf{z})$, with the condition that it is fairly straightforward to sample (that is the reason
why standard normal distributions are often the distribution of choice). Let $\mathbf{X}=f(\mathbf{Z})$
be a transformed variable, with a function $f$ that is invertible. If this condition holds, then
$\mathbf{Z}=g(\mathbf{X})$, where $g=f^{-1}$. The change of variables formula
states that the probability distribution of the transformed variable is given by:
\begin{equation}
    q_\mathbf{X}(\mathbf{x}) = q_\mathbf{Z}(g(\mathbf{x})) 
    \left| \mathrm{det} \left( \frac{\partial g(\mathbf{x})}{\partial \mathbf{x}} \right) \right|.
\end{equation}
The term $\partial g(\mathbf{x}) / \partial \mathbf{x}$ is the Jacobian matrix and takes into
account the change of probability volume during the transformation. Its role is to
force the resulting distribution to be a proper probability distribution with unit
integrated probability. Since the transformation is invertible, the equality 
$\partial g(\mathbf{x}) / \partial \mathbf{x}=(\partial f(\mathbf{z}) / \partial \mathbf{z})^{-1}$ holds, so that
one can rewrite the previous expression as:
\begin{equation}
    q_\mathbf{X}(\mathbf{x}) = q_\mathbf{Z}(\mathbf{z}) 
    \left| \mathrm{det} \left( \frac{\partial f(\mathbf{z})}{\partial \mathbf{z}} \right) \right|^{-1}.
\end{equation}

Designing invertible transformation that can be trained to produce generative models over 
complex datasets is difficult. For this reason, normalizing flows make use of the fact that
the composition of invertible transformations is also invertible. Then, if $f=f_M \circ f_{M-1} \circ \cdots \circ f_1$, the
transformed distribution is
\begin{equation}
    q_\mathbf{X}(\mathbf{x}) = q_\mathbf{Z}(\mathbf{z}) 
    \prod_{i=1}^M \left| \mathrm{det} \left( \frac{\partial f_i(\mathbf{y_i})}{\partial \mathbf{y_i}} \right) \right|^{-1},
    \label{eq:flow}
\end{equation}
where $\mathbf{y}_i=f_{i-1} \circ \cdots \circ f_1(\mathbf{z})$ and $\mathbf{y}_1=\mathbf{z}$.
Compositional invertible transformations have made it possible to define
very flexible normalizing flows through the use of deep neural networks. Arguably the simplest invertible transformation
is a linear operation, which can introduce correlation among dimensions. However, they
are not expressive enough. Perhaps the most successful approach to 
define expressive invertible transformations is via coupling layers \citep{Dinh2014}. The idea rests on 
randomly splitting the $d$-dimensional variable $\mathbf{z}$ at each step of the flow in two disjoint sets, $\mathbf{z}^A$ 
of dimension $p$ and $\mathbf{z}^B$ of dimension $d-p$, so that $\mathbf{z}=(\mathbf{z}^A,\mathbf{z}^B)$.
The transformation is defined as
\begin{align}
    \mathbf{y}^A &= h(\mathbf{z}^A; s(\mathbf{z}^B)) \nonumber \\
    \mathbf{y}^B &= \mathbf{z}^B,
\label{eq:coupling}
\end{align}
where $s(\mathbf{z}^B)$ is, in general, any arbitrary transformation of $\mathbf{z}^B$, while
$h$ is the coupling function. Equation (\ref{eq:coupling})
can be trivially inverted if $h$ is invertible, and the ensuing Jacobian of the transformation
is simply the Jacobian of $h$. The power of coupling resides in the complexity of $s$, which
can be arbitrary, so that complex deep neural networks can be leveraged to produce
very flexible normalizing flows.

We use the rational quadratic neural spline flows of \cite{Durkan2019} as our
coupling functions of choice. They have been
demonstrated to produce very good results in approximating high-dimensional complex
probability distributions. In this approach, the coupling functions are monotone rational
quadratic splines, where the knots defining the splines are obtained by applying a 
neural network to $\mathbf{z}^B$.

Our specific normalizing flow model consists of the sequential application of 5 steps,
to allow for a more expressive posterior distribution modeling. Each step contains 
a linear transformation using an LU decomposition \citep{glow18} and a neural spline coupling 
transformation. The neural network $s$ is
a fully connected residual network \citep{He2015} with a hidden dimensionality of 256 and
rectified linear units \citep[ReLU;][]{relu10} as activation functions. A dropout of 0.4 is 
applied to prevent any kind of overfitting.

\begin{figure*}
    \centering
    \includegraphics[width=0.7\textwidth]{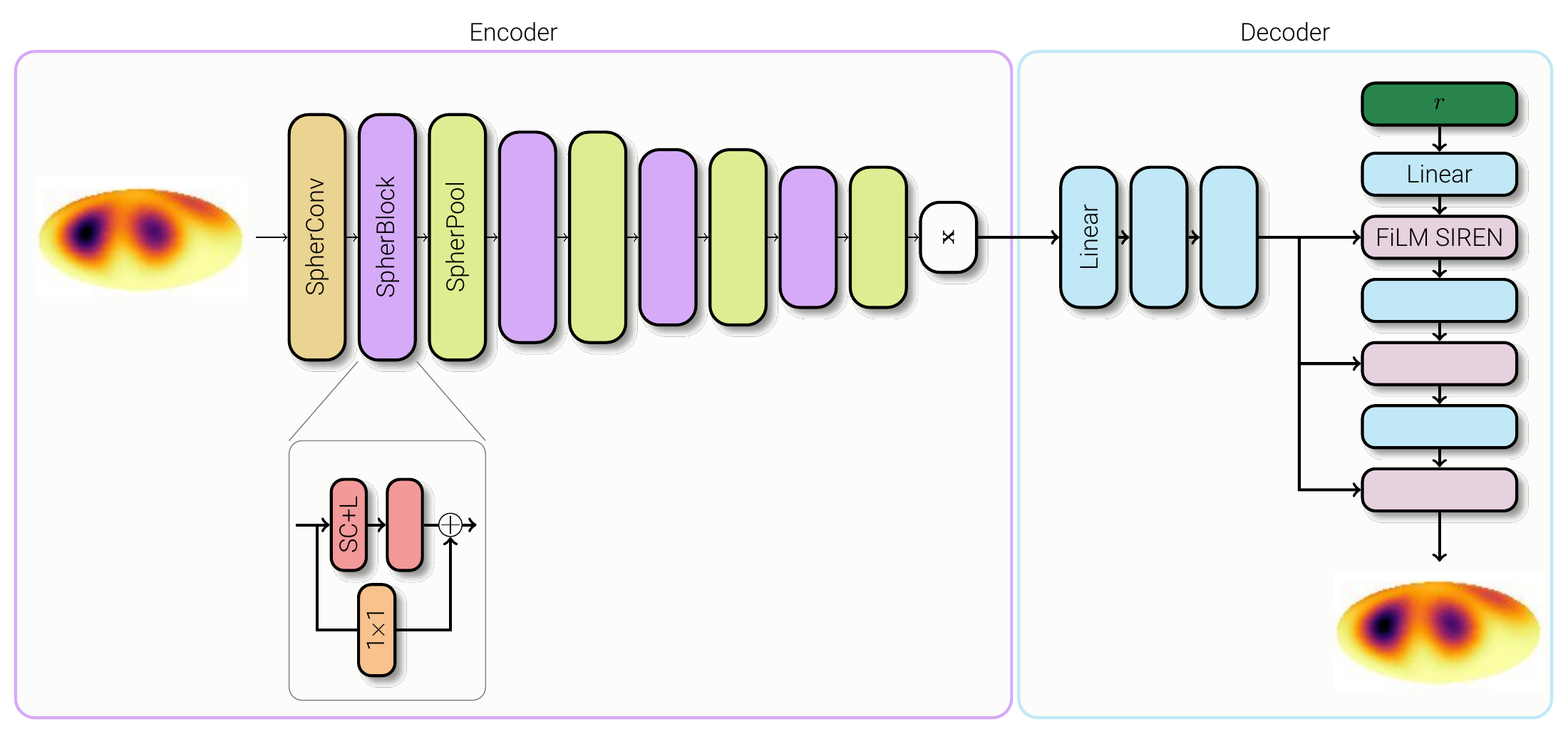}
    \caption{Autoencoder used for the compression of stellar temperature surface maps in
    a latent representation of reduced dimensionality. The encoder (violet block) uses spherical 
    convolutions and pooling in the HEALPix representation. The decoder (blue block) uses 
    modulated SIREN layers, described by Eqs. (\ref{eq:siren}) and (\ref{eq:siren_film}).}
    \label{fig:autoencoder}
\end{figure*}

Since we want to model the posterior distribution $p(\mathbf{T}|D)$, we need to
take into account the conditioning on the observed data. Conditional generative models 
can be modeled with normalizing flows straightforwardly by conditioning all previous 
probability distributions on a latent vector. We discuss how a latent vector 
can be extracted from the observations for conditioning in Sec. \ref{sec:transformer}. 
From a practical point of view, it is enough to concatenate the conditioning vector $\mathbf{o}$
to the input of the residual network, so that one uses $s([\mathbf{z}^B,\mathbf{o}])$
instead of simply $s(\mathbf{z}^B)$.

\subsection{Autoencoder}
\label{sec:autoencoder}
Although normalizing flows have been showed to do a good job even with very high-dimensional
spaces (ours is of dimension 3072), we have found that reducing the dimensionality produces much better predictions
and the training of the model is greatly accelerated. For this reason, we employ
an autoencoder to compress the surface temperatures in the training set to a 
latent space of dimension 64, with a reduction of a factor 48. An autoencoder can be seen
as a nonlinear extension of the linear principal component analysis \cite[PCA;][]{pearson01}. It
is built by putting together an encoder and a decoder, with a bottleneck between the two.
The encoder neural network ($E$) encodes the input (stellar surface temperature maps)
into a latent vector of reduced dimensionality. The decoder neural network ($D$) then tries to reproduce the output again from the
latent vector. The autoencoder is trained with the 100k stellar surface temperature maps by 
optimizing $|\mathbf{T} - D(E(\mathbf{T}))|^2$, i.e., the mean squared error loss function between 
the input and the output. A schematic representation of the autoencoder is shown in Fig. \ref{fig:autoencoder}.

The encoder is a convolutional neural network. Since we are 
dealing with temperatures in the surface of a sphere, we take advantage of the 
specially simple definition in the surface
of a sphere developed by \cite{2019A&A...628A.129K}\footnote{We use our \texttt{PyTorch} implementation
found in \texttt{https://github.com/aasensio/sphericalCNN}} for $3 \times 3$ convolutions. This exploits the spatial information 
on the sphere to produce expressive intermediate representations. The encoder starts by encoding the input into
32 channels, which then go through four stages of residual spherical convolutional blocks followed by 
an average pooling. After every pooling, the number of channels is doubled. The residual blocks consist
of two spherical convolutions preceded by leaky rectified linear units (LeakyReLU) and a residual
connection with a learnable kernel of size 1. The latent vector $\mathbf{x}$ is obtained after applying 
a fully connected layer that produces a vector of dimension 64.

The decoder takes advantage of recent ideas on the field of implicit neural representations. Specifically,
we use a SIREN \citep[sinusoidal representation networks;][]{sitzmann2020implicit}, a simple 
multilayer perceptron (MLP) that takes as input the $xyz$ coordinates of a point in the
surface of the star and returns as output the temperature of this point. SIRENs have been
shown to model many complex signals (images, solutions to partial differential equations, \ldots)
with great flexibility and precision. Given that SIRENs can be trivially extended to produce several
outputs, they can be used as decoders in more complex problems like ZDI.
Since we want this network to model all stars
in our training set, we condition the SIREN to the latent vector $\mathbf{x}$ extracted
by the encoder following the approach of \cite{pigan20}. The SIREN 
computes the following function composition for the position on the
surface described by the vector $\mathbf{r}$:
\begin{equation}
    T(\mathbf{r}) = \mathbf{W}_3 \phi_2^\mathbf{x} \circ \phi_1^\mathbf{x} \circ \phi_0^\mathbf{x}(\mathbf{r}),
    \label{eq:siren}
\end{equation}
where we made explicit that each layer of the SIREN is conditioned on $\mathbf{x}$.
This conditioning is done via a feature-wise linear modulation \citep[FiLM;][]{film17}, which incorporates the latent 
information by projecting it, via a simple 3-layer MLP, using scaling, $\vect{\gamma}_i(\mathbf{x})$, and 
bias, $\vect{\beta}_i(\mathbf{x})$, modifiers. The operation of each SIREN layer is then given
by:
\begin{equation}
    \phi_i^\mathbf{x}(\mathbf{y}) = \sin \left( \vect{\gamma}_i(\mathbf{x}) \cdot \left( \mathbf{W}_i \mathbf{y} + 
    \mathbf{b}_i \right) + \vect{\beta}_i(\mathbf{x}) \right).
    \label{eq:siren_film}
\end{equation}
The SIREN works internally using a representation of size 128, and 
a final layer ($\mathbf{W}_3$) produces a single number from this representation. One of the advantages of using
a SIREN as a decoder comes from their inherent interpolation capabilities. Although trained on
data sampled at 3072 HEALPix pixels, one can later reproduce the temperature at any arbitrary position
on the star. As such, one can potentially use the autoencoder as a prior in standard MAP approaches to DI
and use a latent space of low dimensionality during the optimization. 

\begin{figure}[!b]
    \centering
    \includegraphics[width=\columnwidth]{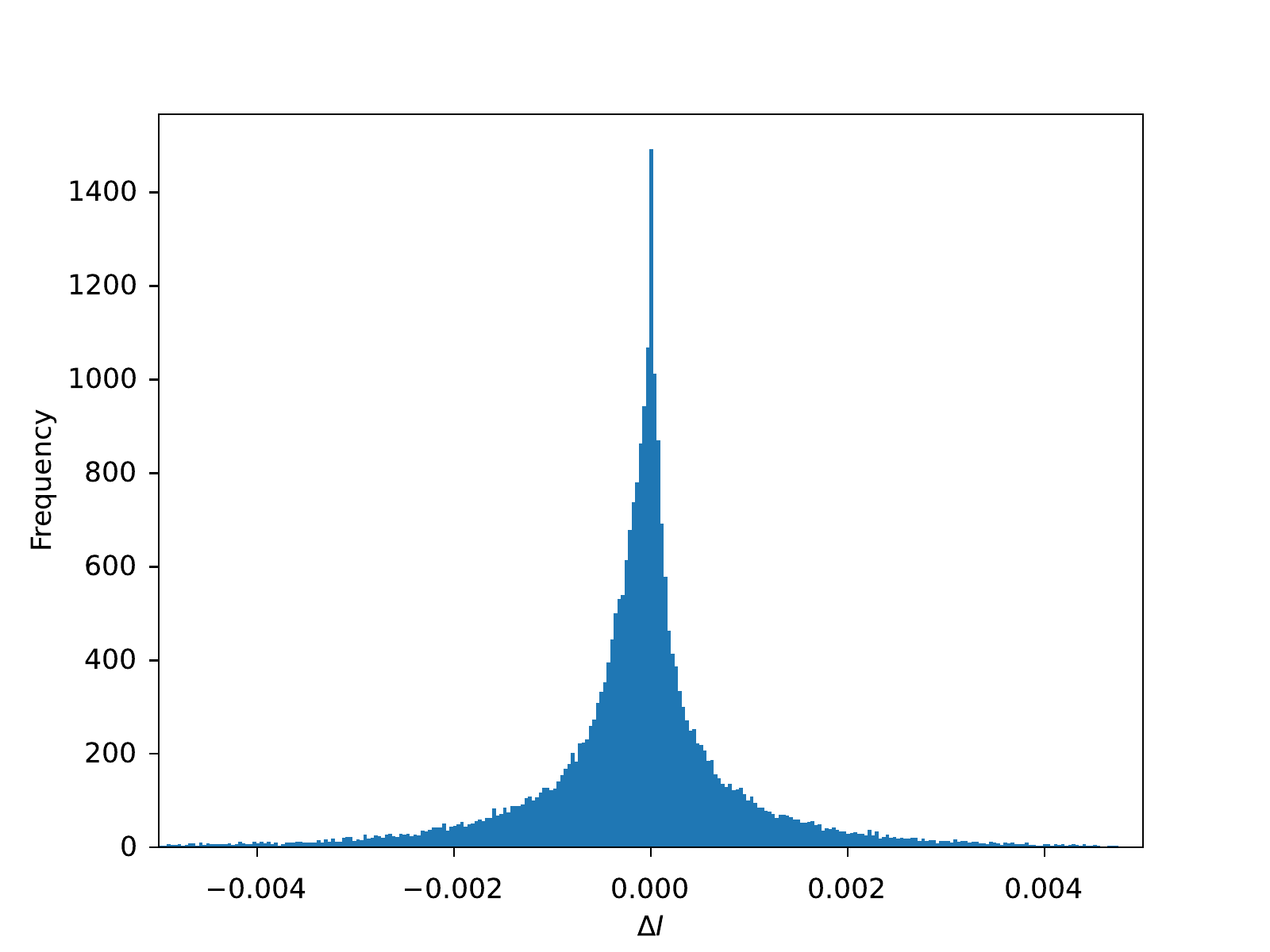}
    \caption{Histogram for $\Delta I$, the difference between the synthetic line profiles in the
    original star and the one produced by the autoencoder.}
    \label{fig:ae_spectra_diff}
\end{figure}

We trained the autoencoder for 200 epochs with a batch size of 128 with the Adam optimizer
\citep{Kingma2014} with a learning rate of $3\cdot10^{-4}$, that is decreased by a factor $1/2$ every 60 epochs.
The number of trainable parameters is 1.547M.
We check for overfitting using 10k stars as validation and find no evidence of it. 
The results show that the dimensionality of the latent space is enough to produce good
representations of the stellar surface, with a root mean square difference of only $\sim$50 K, although
with larger differences in some cases. The difference in the spectra from the original star and 
the one obtained in the star passed through the autoencoder is shown in Fig. \ref{fig:ae_spectra_diff}.
For the majority of cases, the difference is small, with a standard deviation of $\sim$10$^{-3}$. However,
we also find some cases with larger differences, which may have a small impact in the spectral reconstruction.

\subsection{Context: Transformer encoders}
\label{sec:transformer}
\begin{figure}
    \centering
    \includegraphics[width=0.6\columnwidth]{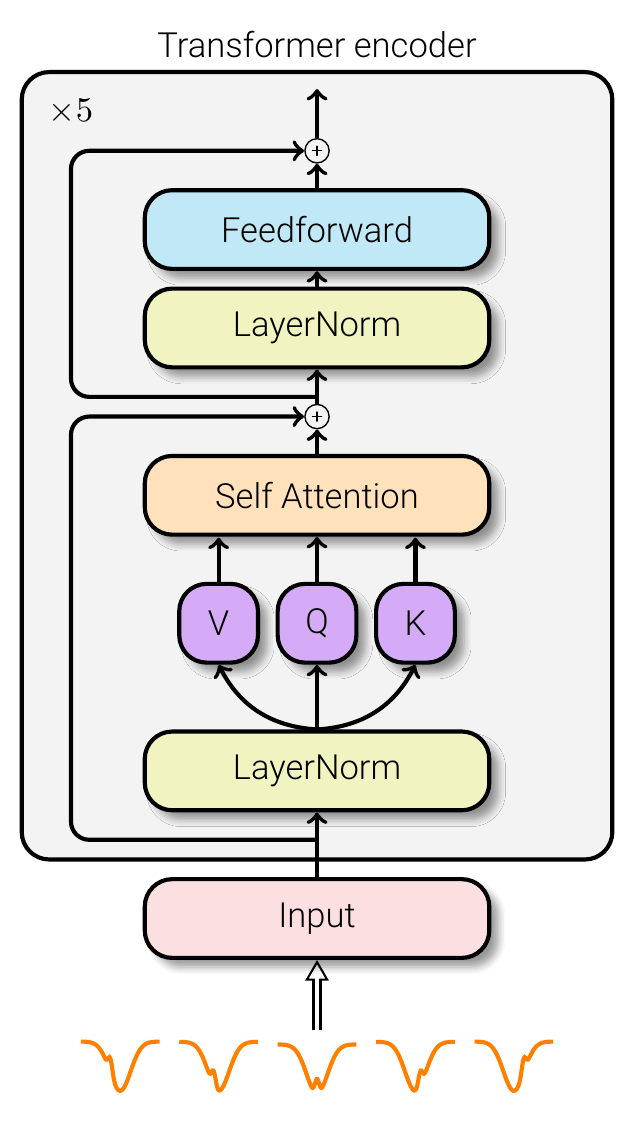}
    \caption{Internal structure of the Transformer encoder, based on self-attention, to deal
    with arbitrary number of phases in time and samples in the spectral direction.}
    \label{fig:transformer}
\end{figure}

Conditioning the normalizing flow on the observations is done via a specific encoder, which
summarizes the observations into a vector of dimension 256. The encoder we need is non-trivial
because it has to deal with two unknowns. First, although we fixed the spectral regions of interest, 
the number of samples in the spectral direction can change from one observation to another. This is a minor 
inconvenience because one can always apply interpolation to sample all observations
in the same wavelength or velocity axis. The second unknown is the
number of observed phases and their specific values, which cannot be fixed a priori, so
it is fundamental to deal with it. Our proposal is to use
Transformer encoders \citep{Transformer17}. Transformers can transform a sequence of arbitrary
length into a sequence of the same length of features of arbitrary dimensionality using self-attention.
The architecture used in this work is displayed in Fig. \ref{fig:transformer}.

The input to the wavelength Transformer encoder for each phase consist of a sequence of $N_\lambda$
wavelength points containing three features. The first one is a quantity proportional
to the wavelength. In our case, we simply use a monotonic vector in the range $(0,1)$ in the
three spectral regions considered, ignoring the wavelength jump between regions. We anticipate
that more elaborate features can be also implemented. The second feature is simply the mean
spectrum computed for all observed phases, which is obviously the same for all phases.
The last one is the residual at each phase with respect to the mean. We concatenate these
three features for all phases and wavelengths of a given batch during training, forming a 
tensor of dimensions $(B \cdot T, N_\lambda, 3)$, where $B$ is the batch size and $T$ is the 
largest sequence length in the batch (we apply binary masks 
to use the appropriate length for each element of the batch). A linear layer
produces feature tensor of size $(B \cdot T, N_\lambda, 256)$. 
A layer normalization layer \citep{ba2016layer} that helps convergence is then applied to the input data. 
Note that we follow the most recent practice of normalizing before entering the 
learnable layers, contrary to the original work of \citet{Transformer17}, which apply
normalizations after the learnable layers.
Matrices of values ($\mathbf{V}$), queries ($\mathbf{Q}$) and keys ($\mathbf{K}$) are built 
by multiplying the inputs for
the whole sequence with trainable matrices. The self-attention layer then computes
the following operation:
\begin{equation}
    \mathrm{Att}(\mathbf{Q},\mathbf{K},\mathbf{V}) = 
    \mathrm{softmax} \left( \frac{\mathbf{Q} \mathbf{K}^T}{\sqrt{d_k}} \right) \mathbf{V},
\end{equation}
where $d_k$ is the dimensionality of the queries and keys, which we set to 256 in our case.
The product of the queries and keys matrices gives as a result a score matrix. This
matrix gives information of how much focus one element of the sequence has to put
on other elements of the sequence, which then shares information of the whole
sequence. This score matrix is then scaled down to allow for more stable
gradients, and a softmax is applied to transform the scores into
probabilities. Finally, these attention weights are applied to the values, normalized
again and passed through a two-layer fully connected network with ReLU activation function. 
Note that the model also contains two residual connections
that accelerate training by reducing gradient vanishing effects. This Transformer layer
is repeated five times in our case. We additionally use two heads (two such encoders
computed in parallel) that will ideally focus on different parts of the sequence, and their
outputs are concatenated at the output. The resulting output of the
Transformer encoder has, therefore, dimensions $(B \cdot T, N_\lambda, 256)$, where all elements of the 
output sequence have attended to all elements of the input sequence, as explained above.

We then average over the spectral direction and reshape the tensor to dimensions
$(B, T, 256)$. These tensors can then be considered to encode the necessary spectral
information for every phase of each observation of the batch. It is at this moment
where it is crucial to include information about the specific phase at which each observation is done. We
do it by using a FiLM layer, which linearly modulates the encoding tensor using
a scaling and a bias function obtained from the phase (as a real number between 0 and 1).
A second Transformer encoder following exactly the same description as above
then aggregates the temporal information, giving as
output a tensor of size $(B, T, 256)$. This tensor is averaged over time, resulting in
a tensor of dimensions $(B, 256)$. Finally, we add the information about
the rotation velocity (normalized to 80 km s$^{-1}$) and inclination of the axis of rotation of the star
(we use $\sin i$, which is constrained to be between 0 and 1), again with a FiLM layer. The result
is the context conditioning vector $\mathbf{o}$ of dimension 256, that is then fed to the flow
to estimate the posterior distribution.

\begin{figure*}
    \centering
    \includegraphics[width=0.9\textwidth]{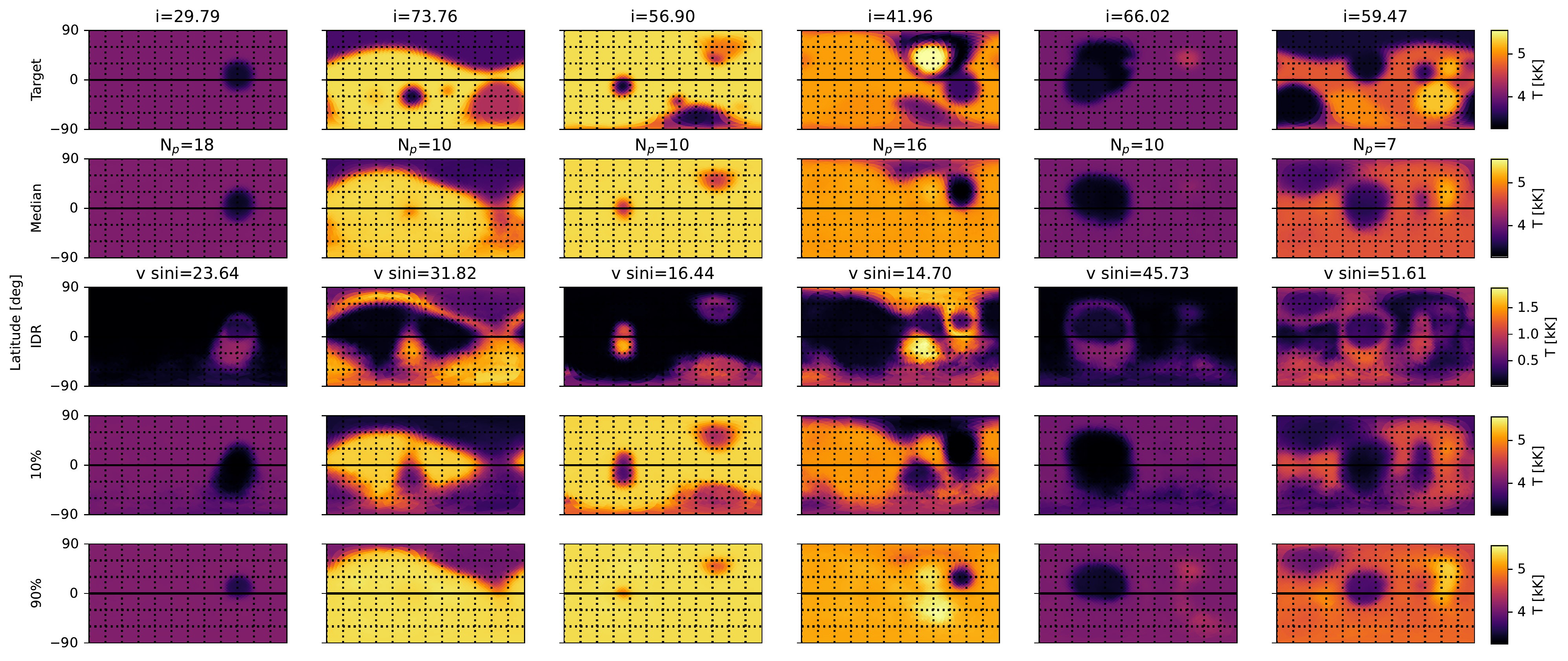}\vspace{0.3cm}
    \includegraphics[width=0.9\textwidth]{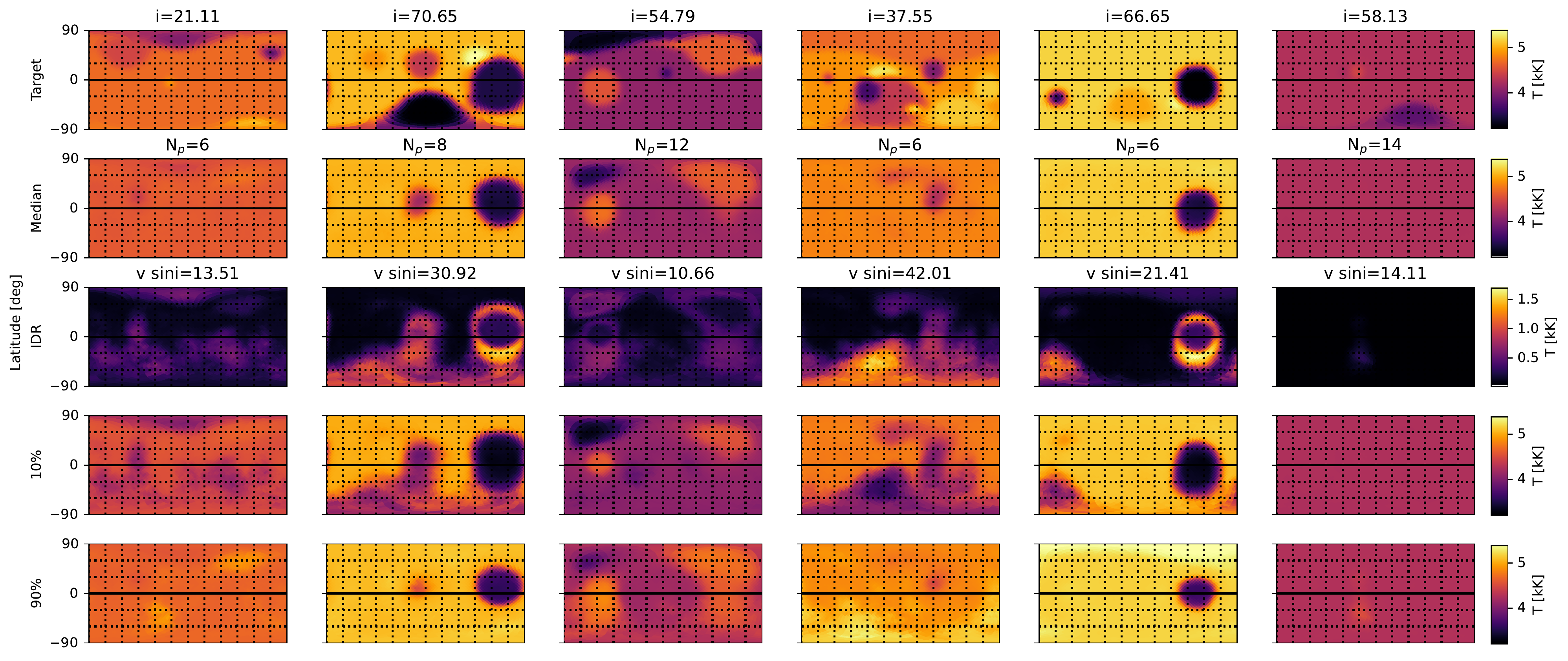}
    \caption{Results of the Bayesian inference for 12 (6 per panel) validation stars. The rotation velocity, number
    of phases and $v \sin i$ of each star is indicated. The first
    row shows the target surface temperature map. The second and third row show the median and
    the interdecile range. The last two rows display the percentiles 10 and 90. All units are given in kK.}
    \label{fig:validation_cases}
\end{figure*}

\subsection{Training}
Using Eq. (\ref{eq:flow}), and conditioning it on the observations, the normalizing flow produces the
following approximation to the posterior (transformed to logarithms):
\begin{equation}
    \log q_\mathbf{X}(\mathbf{x}|\mathbf{o}) = \log q_\mathbf{Z}(\mathbf{z}) +
    \sum_{i=1}^M \log \left| \mathrm{det} \left( \frac{\partial f_i(\mathbf{y_i|\mathbf{o}})}{\partial \mathbf{y_i}} \right) \right|^{-1}.
\end{equation}
Our aim is to train the parameters of the model (that we encode in the vector $\vect{\theta}$) so that the approximate
posterior produces a good reproduction of the real posterior distribution. We do it by minimizing the
expected value of the Kullback-Leibler divergence over all possible observations:
\begin{equation}
    L(\vect{\theta})= \int \mathrm{d}\vect{o} p(\vect{o}) \int \mathrm{d}\mathbf{T}
    p(\mathbf{T}|\mathbf{o}) \left[ \log p(\mathbf{T}|\mathbf{o}) - 
    \log q(\mathbf{T}|\mathbf{o},\vect{\theta}) \right].
\end{equation}
The first term is obviously independent of $\vect{\theta}$ and can be considered
to be constant. Neglecting this term and applying the Bayes theorem, we 
can rewrite the loss function as:
\begin{equation}
    L(\vect{\theta})= C -\int \mathrm{d}\mathbf{T} p(\vect{T}) \int \mathrm{d}\vect{o}
    p(\mathbf{o}|\mathbf{T}) \log q(\mathbf{T}|\mathbf{o},\vect{\theta}),
\end{equation}
where $C$ is independent of $\vect{\theta}$ but hard to compute because we do not
have direct access to the posterior $p(\mathbf{T}|\mathbf{o})$.
Dropping this unimportant constant, this expression is especially convenient because one can perform 
the following Monte Carlo estimation (up to a constant) using a batch of size $B$:
\begin{equation}
    L(\vect{\theta}) \approx -\frac{1}{B} \sum_{i=1}^B \log q(\mathbf{T}_i|\mathbf{o}_i,\vect{\theta}),
    \label{eq:loss}
\end{equation}
where $\mathbf{T}_i \sim p(\mathbf{T})$ and $\mathbf{o}_i \sim p(\mathbf{o}|\mathbf{T}_i)$. In other
words, the Monte Carlo estimation of the loss function simply requires to sample surface
temperatures from the prior distribution and then use this to obtain the 
simulated observation, which includes the noise process (or any necessary latent process as described above).

All neural networks are implemented in \texttt{PyTorch} \citep{PyTorch}. The autoencoder
is pretrained as described in Sec. \ref{sec:autoencoder} and the weights are frozen.
The rest of elements of the architecture (normalizing flow, Transformer encoders and 
FiLM conditioning layers) that
are linked with solid arrows in Fig. \ref{fig:model} are trained end-to-end by
minimizing the loss function of Eq. (\ref{eq:loss}). 
We utilize the \texttt{nflows} library, an implementation of normalizing flows in 
\texttt{PyTorch} developed by \citet{nflows}. This library makes it easy to define
normalizing flows and also have a good implementation of the neural spline flows that
we use in this work. For the training we apply the Adam optimizer for 100 epochs with a learning 
rate of 3$\times$10$^{-4}$, which is decreased by a factor $1/2$ every 60 epochs. 
The training time per epoch in one single RTX 2080 Ti GPU is of the order of 14 minutes, giving a total training 
time of around one day. The number of trainable parameters is 11.3M. Almost 6.5M correspond to the normalizing
flow, with each Transformer encoder having around 2.6M parameters\footnote{A trained model 
and the tools needed for retraining can be found in \texttt{https://github.com/aasensio/bayesDI}.}. 

\begin{figure*}
    \centering
    \includegraphics[width=\textwidth]{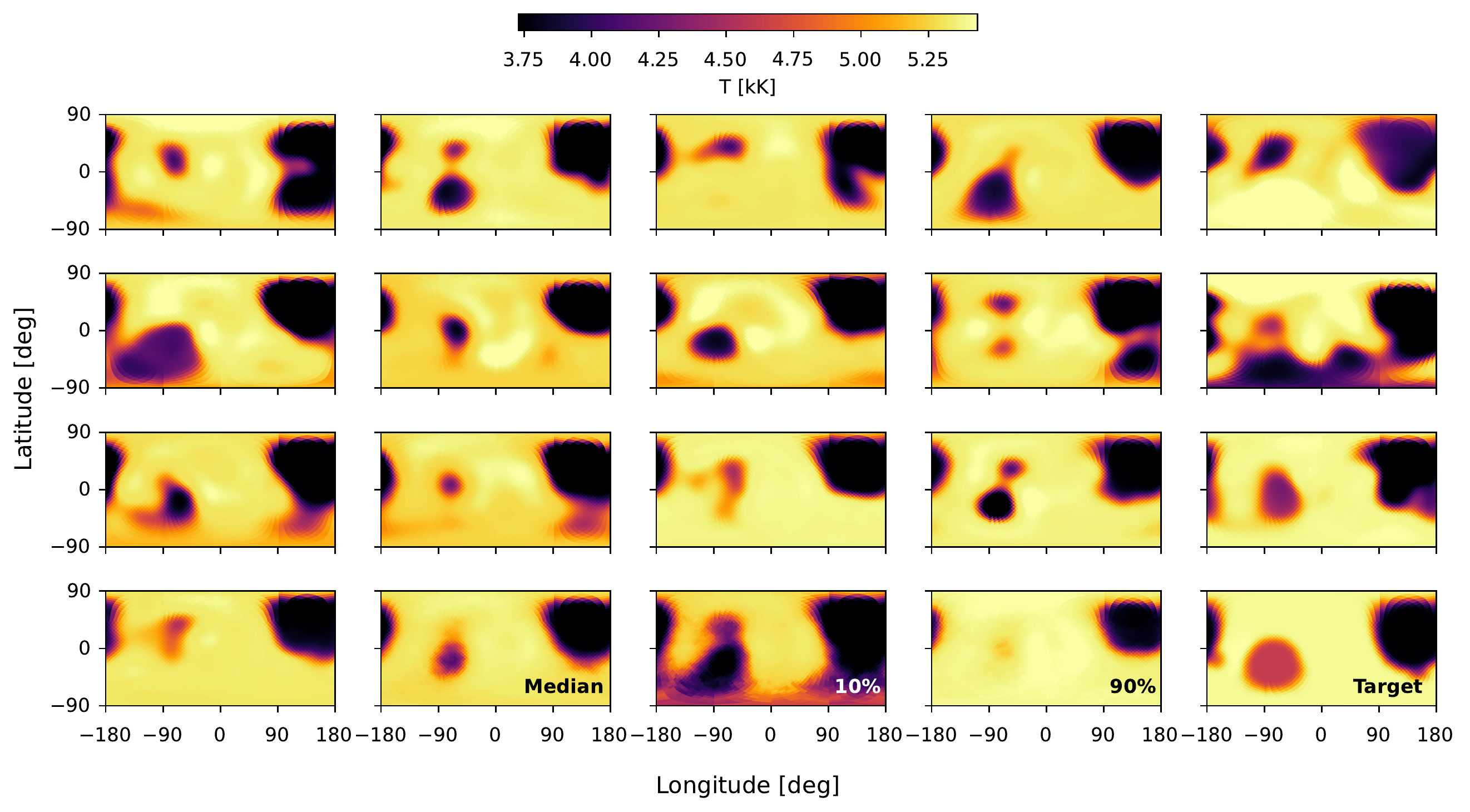}
    \caption{Samples from the posterior distribution, median, percentiles 10 and 90
    and the original target temperature surface map. This case has a rotation velocity of 
    49 km s$^{-1}$ and an inclination angle of 75$^\circ$, and the star has been observed during 9 phases,
    randomly spread over the rotation period.}
    \label{fig:validation_samples}
\end{figure*}

We carried out a brief analysis of the impact of hyperparameters with the 
objective of finding the optimal ones. The baseline model we use in this work 
is definitively large (with more than 13M parameters) but it can be reduced with
a limited impact on the results. We checked that the size of the context vector $\mathbf{o}$
and the hidden dimensionality of the residual network in the normalizing flow
can both be reduced to 128 with reduced impact. The number of attention blocks
in the Transformers can also be reduced to three. We also verified that increasing
the number of steps in the normalizing flow to 15 barely affects the results.

Only 90\% of the available 100k is used for training, and we use the remaining
10\% for validation purposes and check for overtraining. We do not find evidence
of it. The training proceeds in detail as follows. A batch of stars is extracted from the 
training set. This batch contains the observed spectra, the observed phases for
each star, their rotation velocity and inclination of the rotation axis and the
corresponding stellar surface temperature map. The spectra, phases, velocity and inclination
angle are used by the two Transformer encoders and the FiLM layers to produce
the context vector $\mathbf{o}$. At the same time, the surface encoder of the
autoencoder is used to
produce the latent vector $\mathbf{x}$ for each surface temperature. The 
log-posterior of the normalizing flow is computed with the context and latent
vectors. Backpropagation of this log-posterior allows us to modify the
parameters of the normalizing flow and the Transformer encoders and FiLM layers.

Once the model is trained, samples from the posterior are obtained as follows.
One produces the context vector $\mathbf{o}$ from the observations,
uses standard normal noise in the input to the normalizing flow, and computes the
associated latent vector $\mathbf{x}$ by applying the variable
transformation rule learned by the normalizing flow (see Eq. \ref{eq:flow}). Surface temperature maps
can then be obtained by plugging the sampled latent vector into the
decoder of the autoencoder. Since all these operations are very fast, 
thousands of samples from the posterior can be obtained in less than a
second. Although the high-dimensional posterior is the full result of the inference,
we use simple summaries to show
graphical representations of our results. For instance, the marginal distribution of
temperature in pixel $i$ in the surface is obtained by computing the marginalization over
the rest of pixels:
\begin{equation}
    p(T_i|D) = \int \mathrm{d}T_1 \cdots \mathrm{d}T_{i-1} \mathrm{d}T_{i+1} \cdots \mathrm{d}T_N p(\mathbf{T}|D).
\end{equation}
This marginal distribution is simply computed by computing histograms from posterior samples 
for pixel $i$.

\section{Validation}
To validate the model and understand how to interpret the results, we generate
a set of new synthetic temperature surface maps and compute the emergent spectra
at an arbitrary number of observed phases for different values of the rotation
velocity and inclination angle. Figure \ref{fig:validation_cases} shows the results
for 12 different stars (six stars for each of the two panels) randomly sampled
from the validation set. We use a cartesian projection (latitude-longitude) to show the surface. The reconstructed
surfaces are obtained with the decoder of Fig. \ref{fig:autoencoder} for the 
coordinates $\mathbf{r}$ of the HEALPix pixels for $N_\mathrm{side}=16$. 
The upper row in each panel displays the target temperature
distribution on the surface of the star. The second row displays the median (percentile 50) of the posterior
distribution, while the remaining rows try to summarize the uncertainty. The third row
displays the interdecile range (IDR; the range between the percentiles
10 and 90 of the distribution). Given that the posterior can be far from
Gaussian, the IDR is a measure of dispersion that better provides information on
large excursions from the median than the standard deviation (the standard deviation is approximately
given by IDR/2.56 in the Gaussian case). Since the distribution in each pixel does not need to be
symmetric, the fourth and fifth rows show the percentiles 10 and 90, which can
be used to check the expected variability of the temperature in each pixel.

\begin{figure}
    \centering
    \includegraphics[width=\columnwidth]{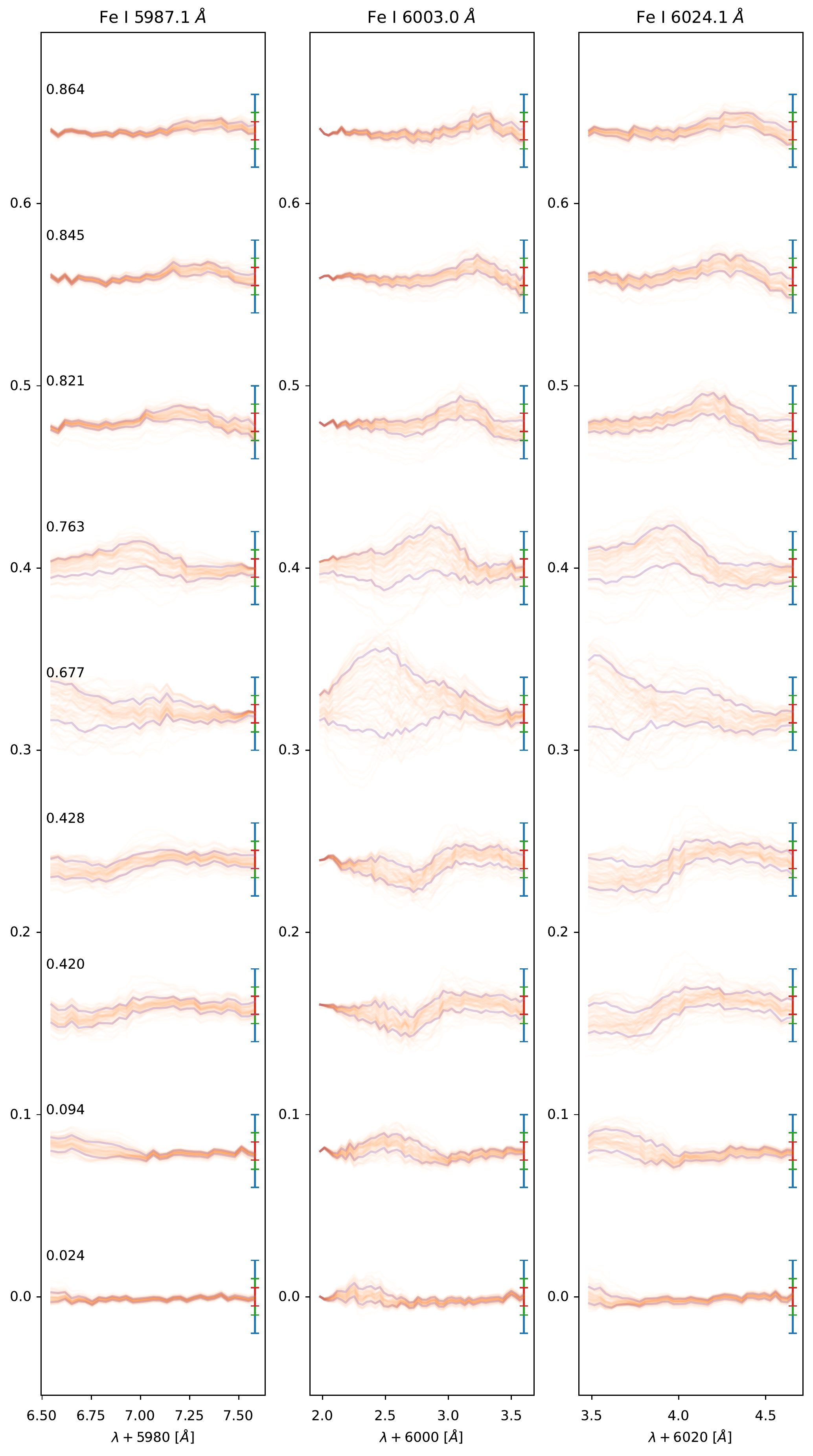}
    \caption{Residual from the synthesis in the posterior samples and the observations for
    the case of Fig. \ref{fig:validation_samples}. Samples are shown in orange, with violet curves
    indicating 68\% interval estimate. The red, green and blue error bars show residuals of 0.02, 0.01,
    and 0.005, respectively.}
    \label{fig:spectra_validation}
\end{figure}

These results show that the median of the distribution is a fairly good
representation of the temperature distribution on the surface of the star. Dark
and bright spots are correctly obtained, although many small-scale structures
are lost and cannot be recovered. Given that the training was done with $i \in [10^\circ,85^\circ]$
to avoid ambiguities, the results are clearly biased towards a better reconstruction of 
the northern hemisphere. However, this does not affect the behavior of the algorithm. 
Despite this bias, we find no artifacts
in the southern hemisphere. Of special relevance are cases in which the rotation axis
is pointing to the observer ($\sin i \sim 0$), for which practically no
information about this hemisphere is encoded in the observations. The absence
of artifacts is a direct consequence of
the regularizing effect of the prior and the marginalization of the Bayesian
inference. We find that the temperature in the southern hemisphere is roughly
similar to the underlying temperature of the star.

Although the position in latitude and size of all inferred structures is correctly recovered
on average, there is some ambiguity, as marked by the bright regions in the IDR. The regions in the
shape of rings (for instance the second case in the lower panel) define that there
is some uncertainty in the specific position of the spot, mainly in the latitude. However, this uncertainty
is smaller than the size of the spot, both in latitude and longitude. The recovery of 
spots is quite robust, both in the regions colder and hotter that the surroundings, as indicated
by the percentiles 10 and 90.

Though the percentiles and the IDR shown in Fig. \ref{fig:validation_cases} are good summaries of the dispersion of 
the posterior distribution, it is also good to see samples from the posterior distribution. An example is shown
in Fig. \ref{fig:validation_samples}, where we display 13 samples from the posterior, together with the
median, percentiles 10 and 90 and the original map (all marked with labels). Samples from the 
posterior all look similar, although spots are placed at different positions in the
surface. The large cold spot and the slightly smaller less cold spot are clearly visible in the median. 
The percentiles indicate that the posterior distribution is fairly asymmetric for the hotter spot, which
is clearly visible in percentile 10 but not barely in percentile 90. All samples that are compatible with the 
observations display the large cold spot although the hotter spot location is more uncertain. The ability of
detecting hotter spots crucially depends on the specific spectral lines selected, which might
not be optimal in our case. The very small-scale spot between both large spots is not visible in the median map, 
fundamentally because its small size is below the typical resolution element of $2v_\mathrm{rot} \sin i/W$
\citep[e.g.,][]{vogt87}, with $W$ the considered Doppler widths of the lines. However, we find hints of it 
in the percentile 10 map.

Figure \ref{fig:spectra_validation} shows the synthetic spectra using samples from the posterior distribution
of Fig. \ref{fig:validation_samples}, also known as posterior predictive checks. 
For convenience, we display the residual between the 
original synthetic observations and each sample in orange curves. Violet curves show 68\% interval estimates. The
error bars act as visual guides indicating amplitudes of 0.02, 0.01, and 0.005 for blue, green and 
red, respectively. It is clear from these results that, although there is no apparent bias on
the inference, the approximated posterior distribution is too broad in some cases. Although
the amplitude of the residuals is clearly close to the noise level of $10^{-3}$ in some phases, much larger
amplitudes are found in other rotation phases. We have thoroughly explored whether this can be solved
by changing the hyperparameters and architecture of the model and we have had little success. Almost all
models we have considered successfully converge towards predicting the correct median but with
an overestimation of the uncertainty. A potential solution to this issue is to radically augment
the training set, something we have not pursued due to computational limitations, but plan to
do in the near future.

\begin{figure*}
    \centering
    \includegraphics[width=\textwidth]{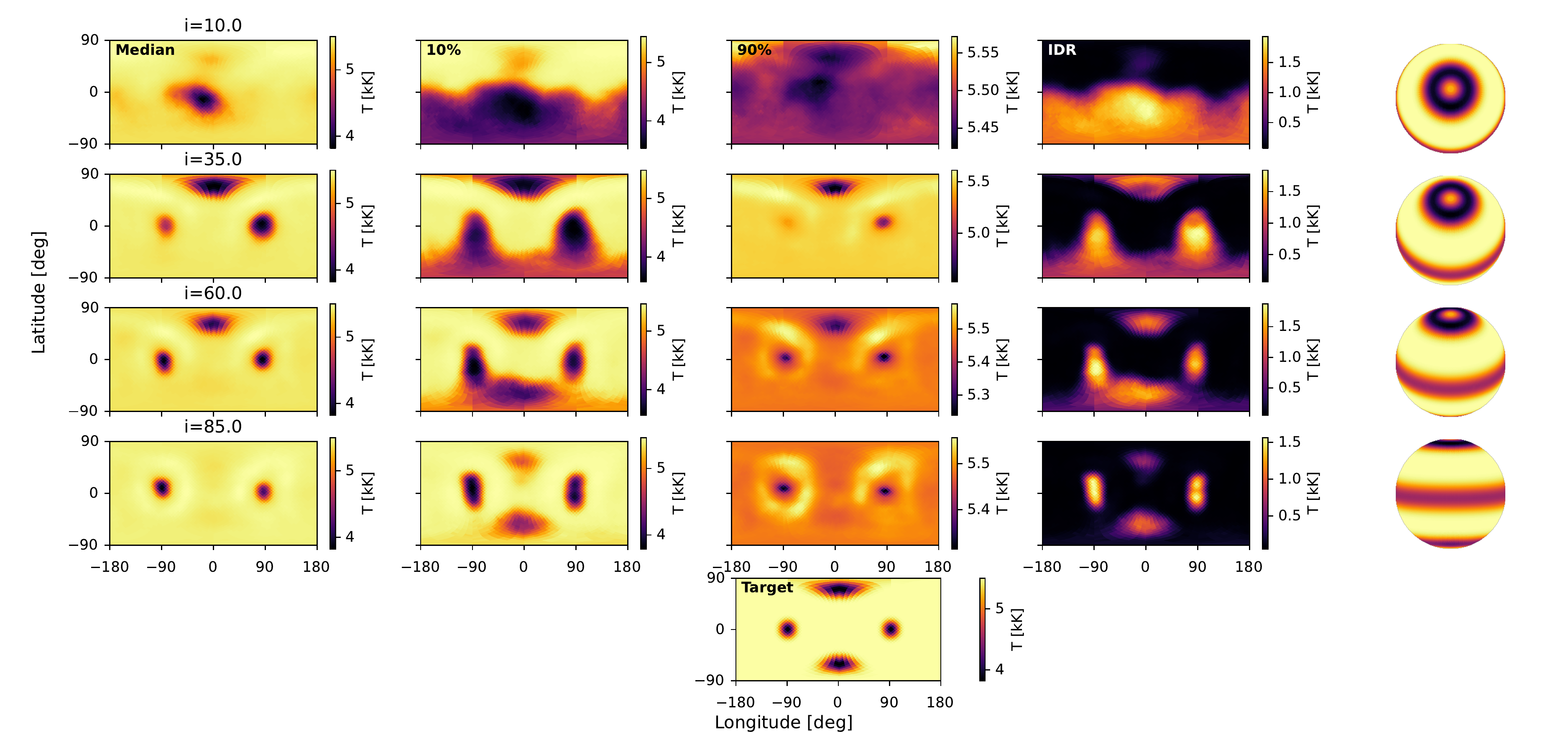}
    \caption{Reconstruction of a synthetic case for different inclinations of the rotation angle. The star
    rotates at 30 km s$^{-1}$ and is observed at 15 equidistant rotation phases. First column: median of the
    the marginal posterior distribution per pixel. Second, third and fourth columns: percentiles 
    10 and 90 of the
    same distribution, and IDR, respectively. Fifth column: pictorial representation of the 
    observability of the active
    regions for each inclination angle. Lower inset: original temperature distribution in the star.}
    \label{fig:variability_sini}
\end{figure*}

As a final validation of the results, we analyze how results behave when the inclination
of the rotation axis changes for the same star. To this end, we display in Fig. \ref{fig:variability_sini} a
tailor-made star of 5500 K with three spots of 3500 K and 0.2 rad (11.5$^\circ$) of radius. 
Two of the spots are located 
at the equator with a longitude difference of 180$^\circ$. The remaining two are located at longitude 0$^\circ$
and at latitudes 70$^\circ$ and $-60^\circ$. The star is assumed to be rotating with an
equatorial velocity of 30 km s$^{-1}$ and
observed at 15 equidistant rotation phases for a whole rotation. The lower panel displays the
target temperature map in cartesian projection. The four panels in the
left column shows the median of the inferred temperature posterior distribution for four values of the
inclination angle. The three central columns show the percentiles 10 and 90 and the IDR, respectively. 
The rightmost column is a simple pictorial representation of the visibility of the active regions
with rotation, obtained by simply averaging the temperature over many rotations. It shows the relative 
position of the dark spots with respect to the observer.

The spot located close to the north pole of the star is captured for observations $i=10^\circ$, $35^\circ$
and $i=60^\circ$. The Doppler effect for this spot for observations at $i=10^\circ$ is weak because it is almost
on the plane of the sky, but the spot is correctly located. For observations at $i=85^\circ$, although the 
Doppler effect is maximum, the projected area on the plane of the sky is minimum and the results show 
only a hint of its detection in the lowest percentile.
The spot in the southern hemisphere barely appears in the median maps and it is only
clearly found in the percentile 10\% maps. The equatorial spots are very nicely detected in the
median maps and in both percentiles, although there are large uncertainties on the precise position
of the spot in the percentile 10\% map. The specific location of these spots for
an observation at $i=10^\circ$ is clearly wrong, though. Such poor reconstruction is
not surprising since the projected velocity is $v_\mathrm{rot} \sin i \approx 5$ km s$^{-1}$.
The percentiles indicate some confusion in the inferred
position of these spots and the one in the southern hemisphere. Interestingly, we witness the appearance 
of a bias towards large temperatures in the areas surrounding the dark spots. These structures are also
seen in some of the examples of Fig. \ref{fig:validation_cases}. This is a consequence of the 
uncertainty in the exact location in longitude of the spots.

\begin{figure*}
    \centering
    \includegraphics[width=0.98\columnwidth]{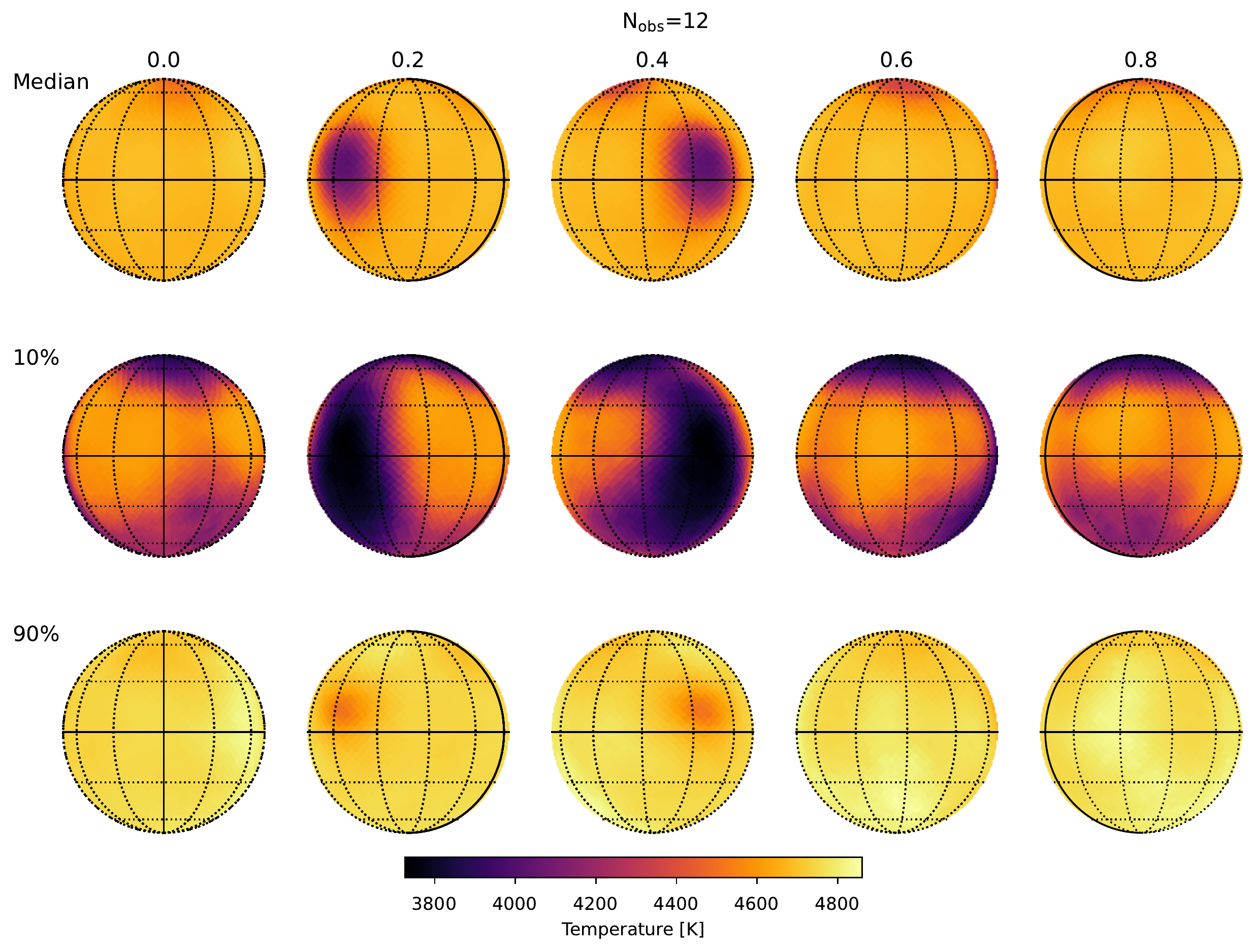}
    \includegraphics[width=0.98\columnwidth]{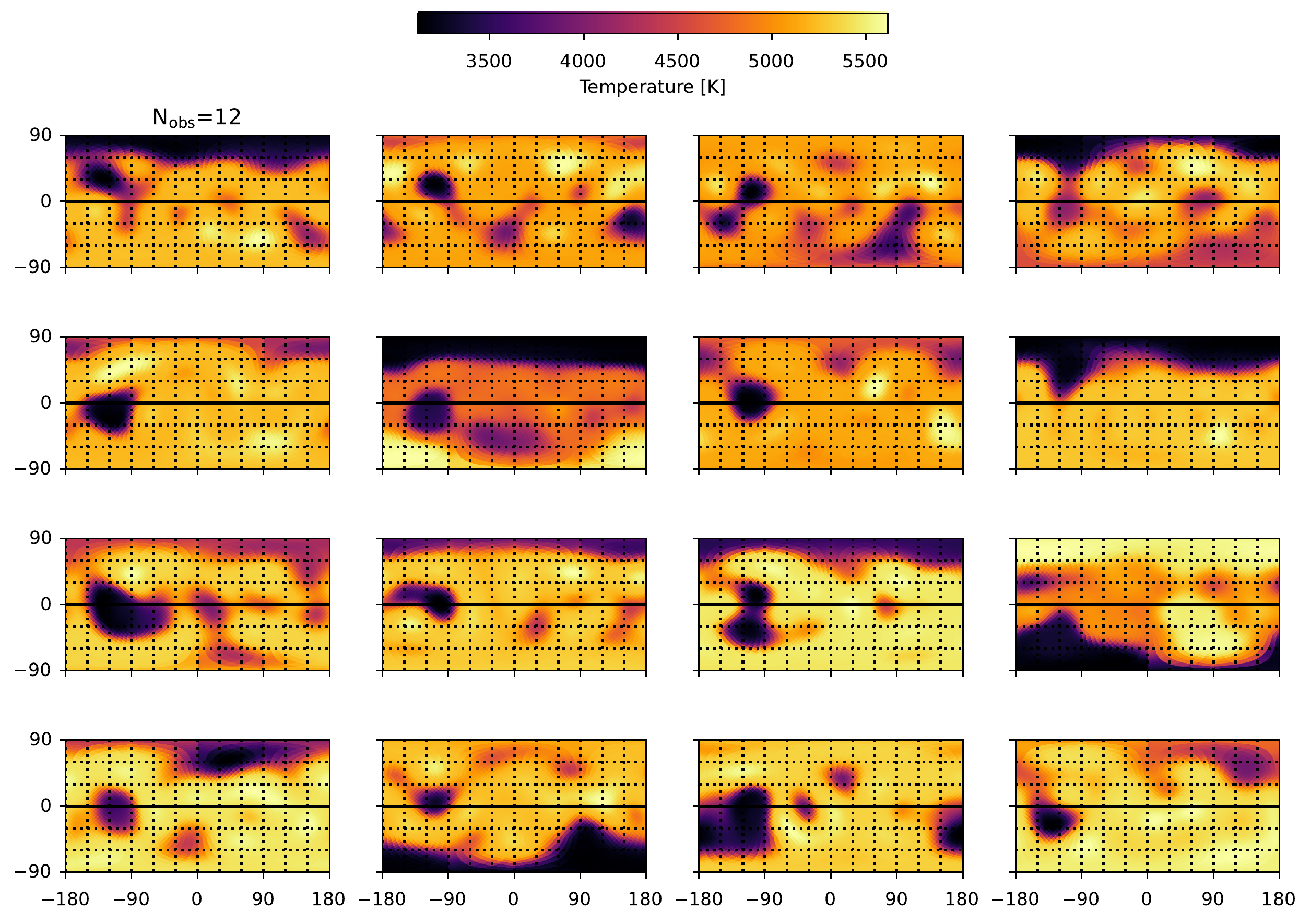}
    \includegraphics[width=0.98\columnwidth]{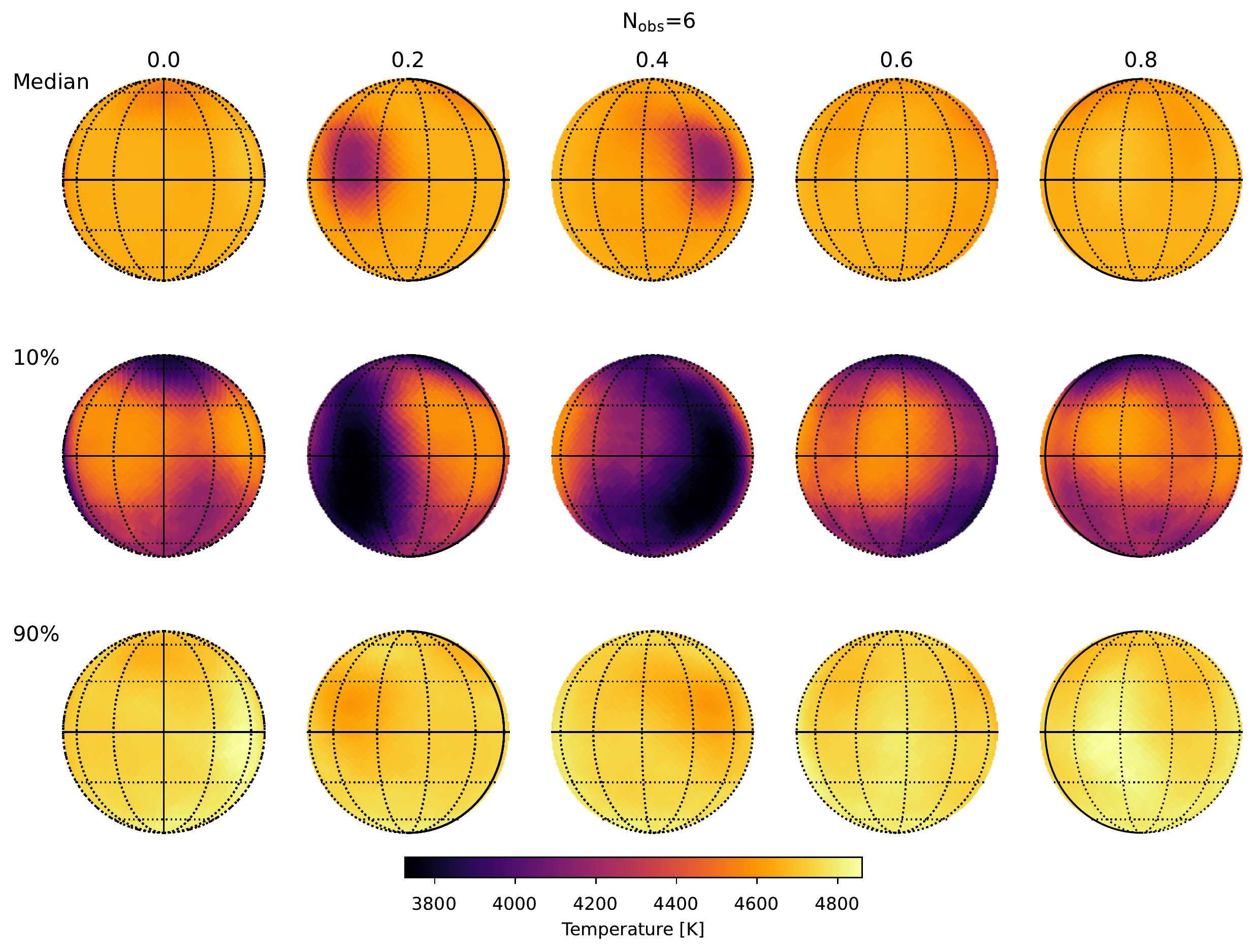}
    \includegraphics[width=0.98\columnwidth]{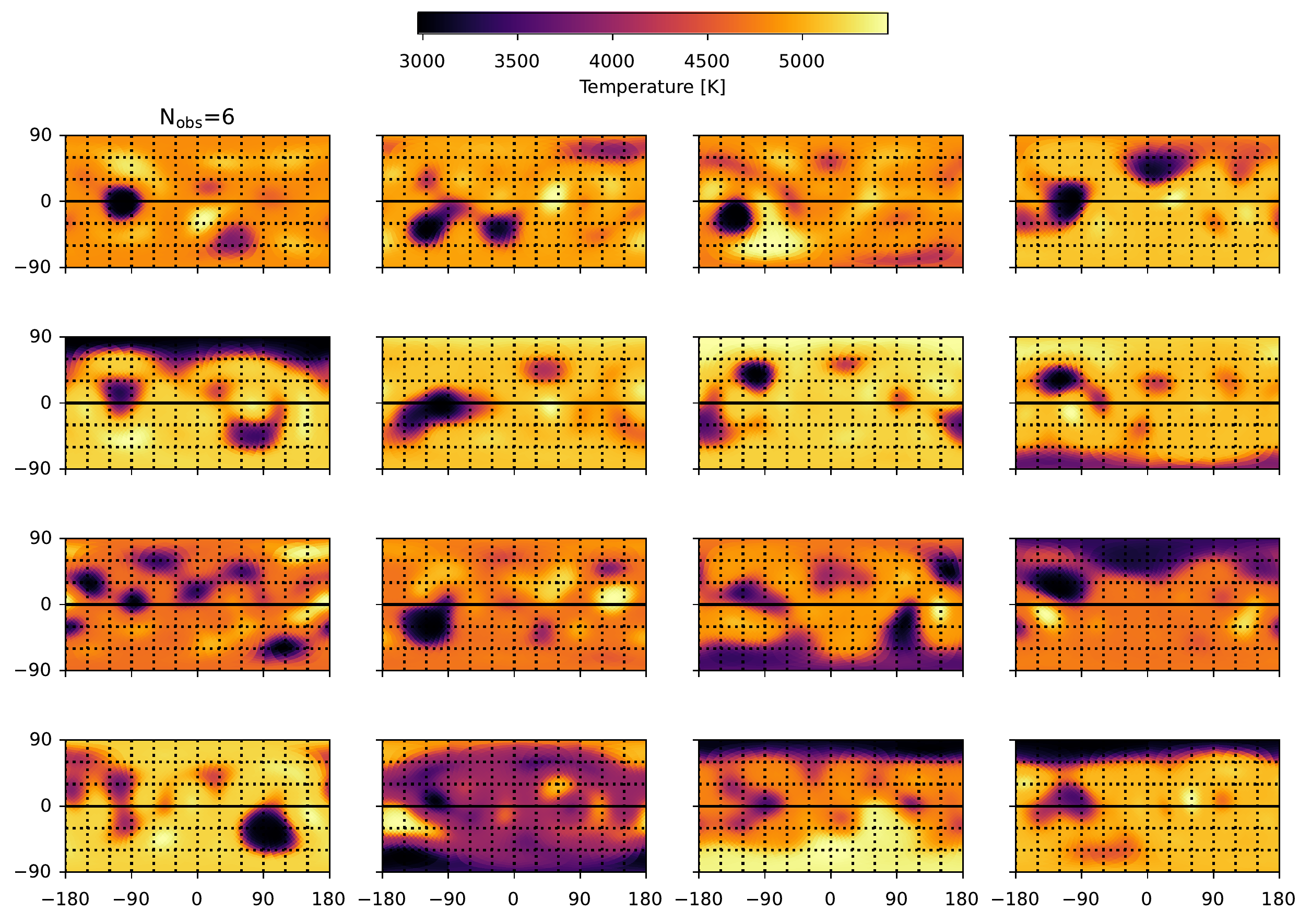}
    \includegraphics[width=0.98\columnwidth]{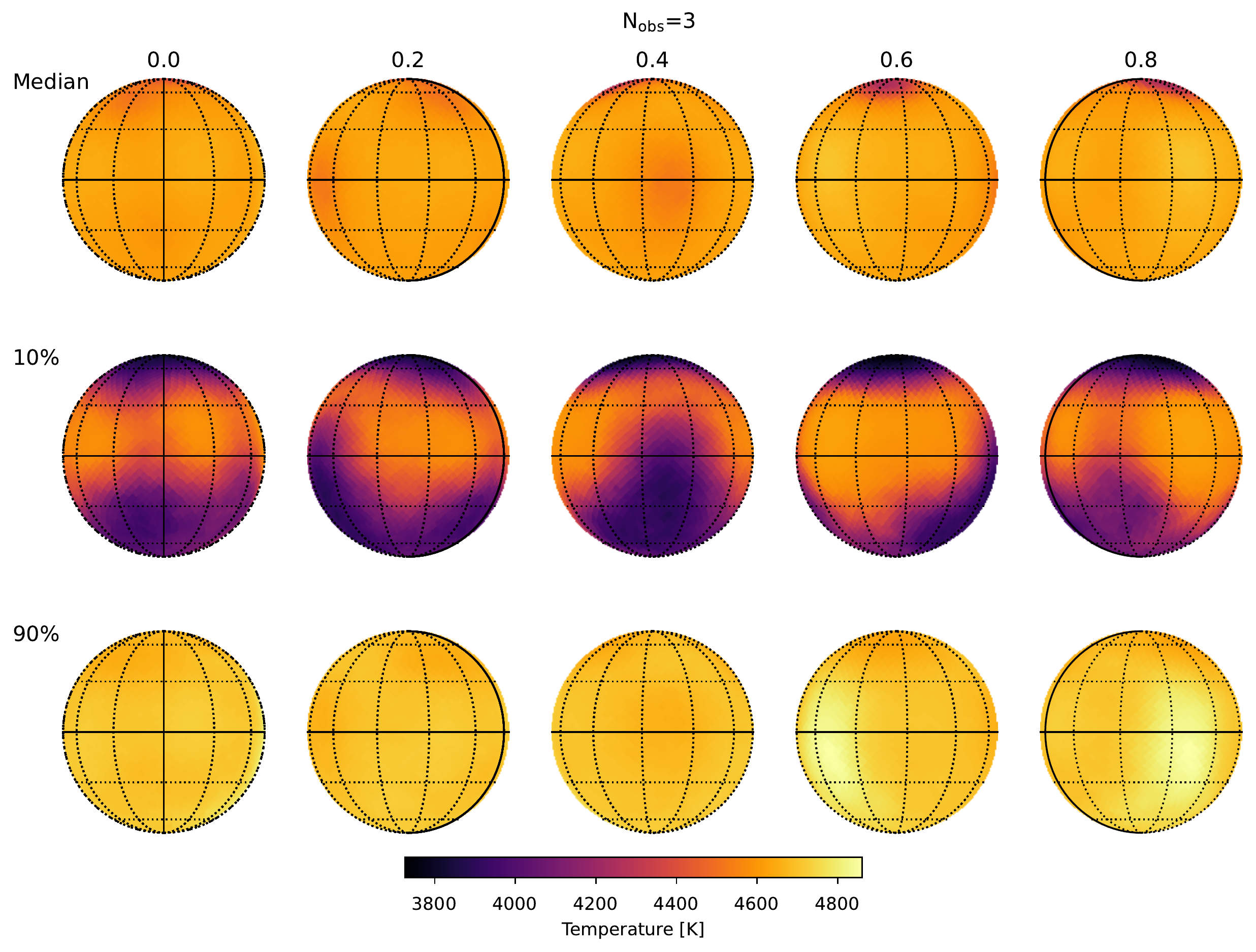}
    \includegraphics[width=0.98\columnwidth]{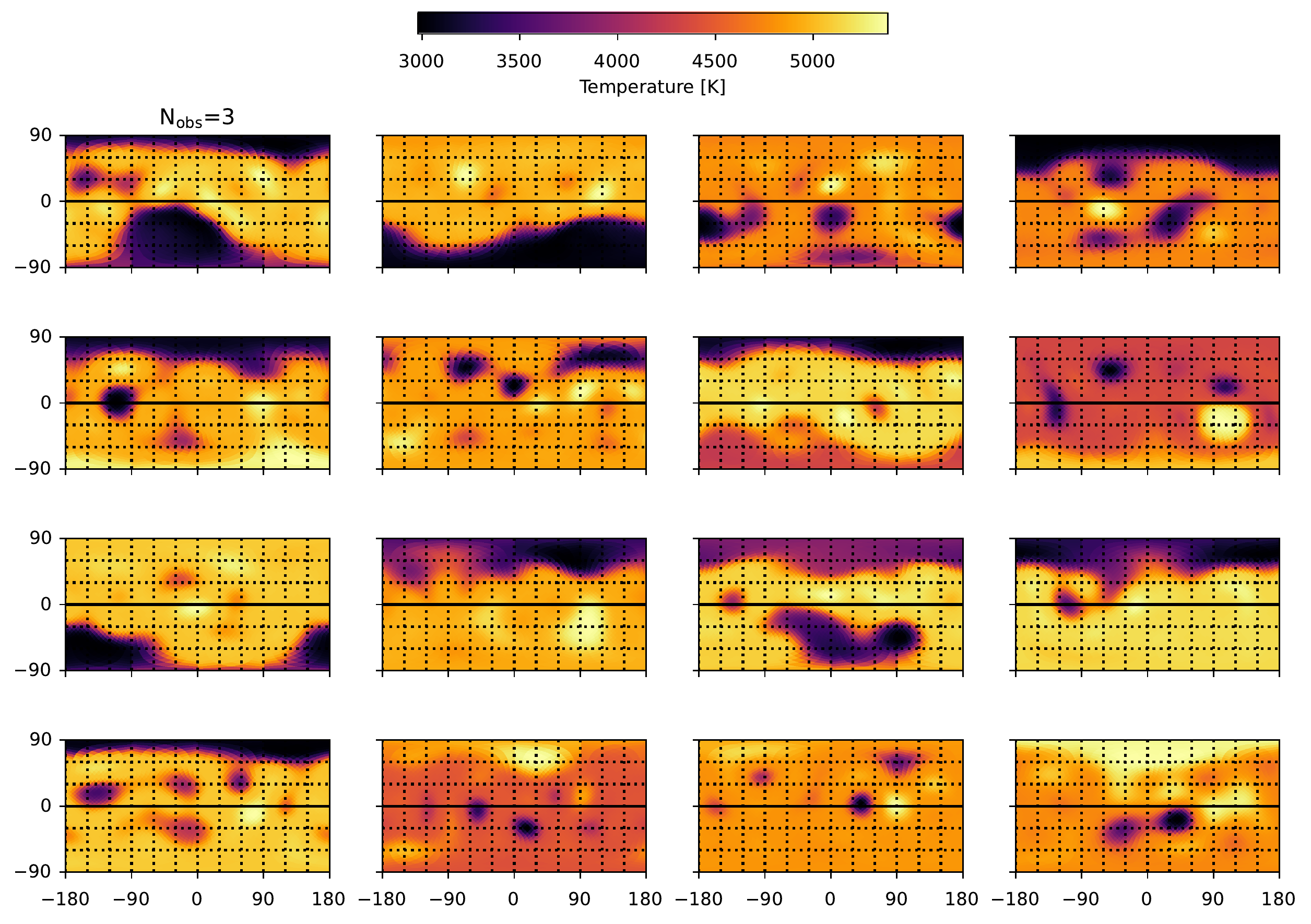}
    \caption{Left panels: median and percentiles 10 and 90 of the posterior distribution for the
    temperature surface map. The star is shown from 5 points of view to have access to
    the inference in all points of the surface. The upper panel displays the results when all
    available observations are used. The middle panel displays what happens when only
    half of the observations are used. The lower panel shows the results when only 3 observations
    are used. Right panels: sixteen posterior samples for each case.}
    \label{fig:iipeg_summary}
\end{figure*}

\section{Application to II Peg}
\label{sec:iipeg}
With the method validated with synthetic observations, we apply it to
observations of the moderately fast rotator II Peg. This star is the primary component in the very active and well-studied RS CVn binary. The observations used here were
presented in \citet{rosen15} but we give a brief summary for completeness.
The observations consist of a total of 12 rotational phases, observed between 2013 June 15 
and 2013 July 1. The observations were performed with the CFHT using
the ESPaDOnS spectropolarimeter. Although the spectra covers the region 3700-10500 \AA\ and
the full Stokes vector is available, we only focus on the three spectral regions discussed in 
Sect. \ref{sec:stars} and on Stokes $I$. The spectral resolution is $R=65,000$. The spectra were reduced 
with the standard Libre-ESpRIT package \citep{donati97}. The resulting SNR per velocity bin
is of the order of 1000. An inclination angle of $i=60^\circ$ and a projected rotation velocity 
of $v \sin i=23$ km s$^{-1}$ is assumed for II Peg \citep{rosen15}. We follow \citet{rosen15}
and add a macroturbulent velocity of 4 km s$^{-1}$ to the spectra to match the
stellar parameters of II Peg. However, given the significant rotation velocity of II Peg, we model it 
by simply reducing the spectral resolution from $R=65,000$ to $R=49,120$.

\begin{figure*}
    \centering
    \includegraphics[width=0.90\columnwidth]{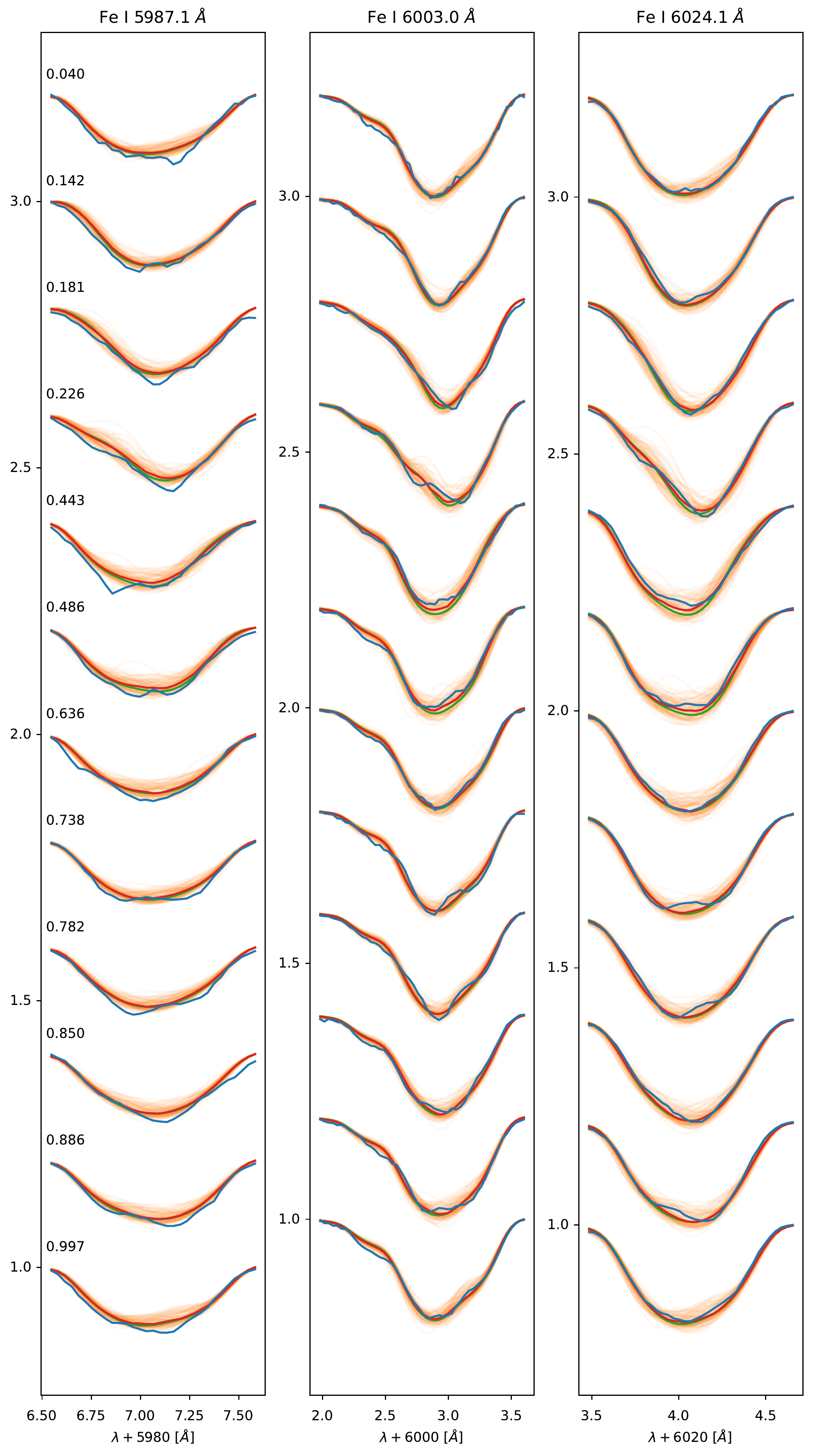}\hspace{1.cm}%
    \includegraphics[width=0.90\columnwidth]{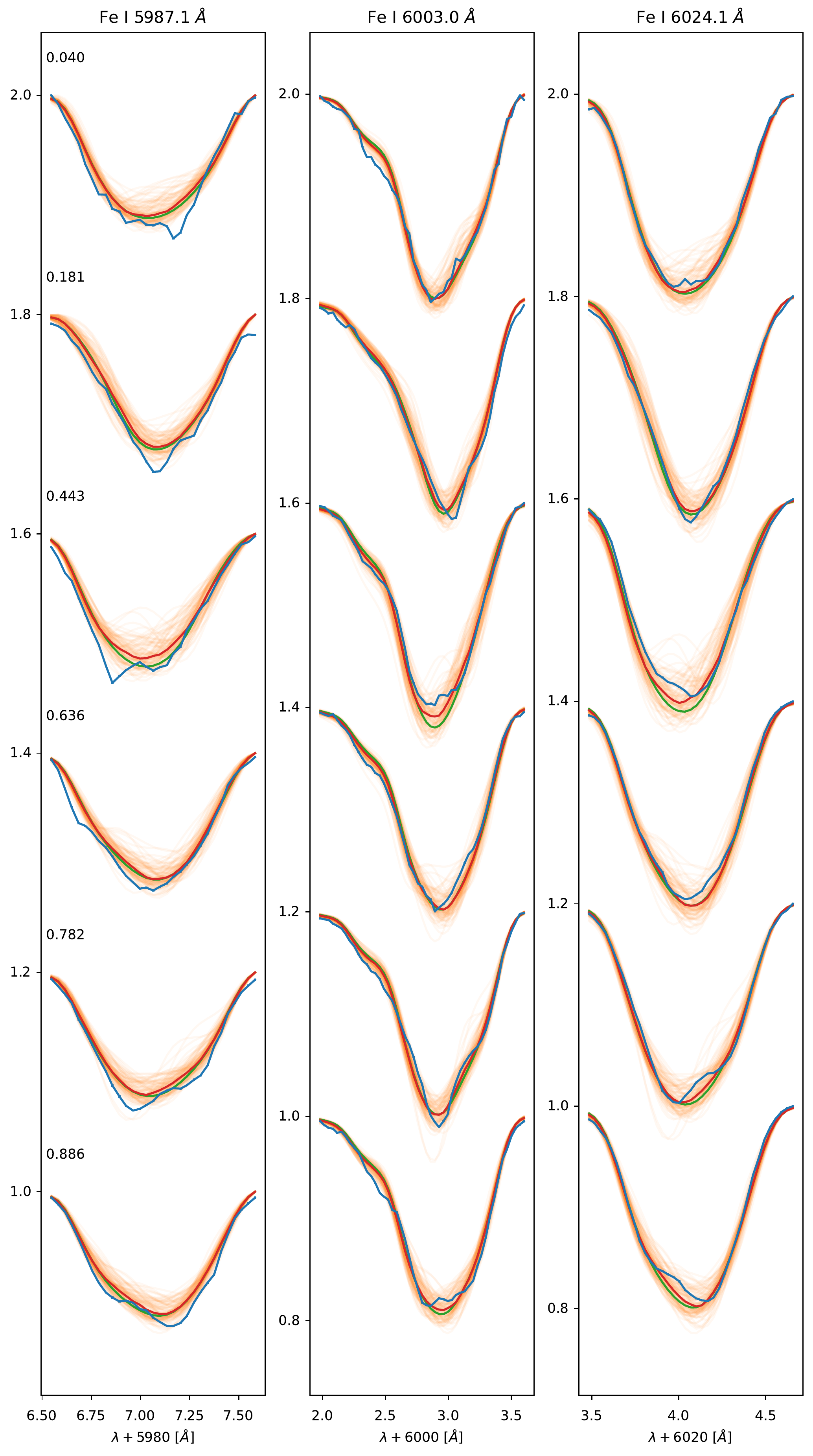}
    \caption{Orange lines: synthesis carried out with samples from the posterior. Blue lines: 
    original observations. Green line: synthesis using the temperatures of the median estimation
    of the posterior. Red line: median of all spectral lines synthesized. The left panel shows the results
    for all phases, while the right panel shows the results when half the observations are used. We show
    the three spectral regions used for the inference, marking with labels the specific rotation phases.
    The lines of each rotation phase are displaced by 0.2 units to avoid crowding.}
    \label{fig:iipeg_spectra}
\end{figure*}

Posterior samples are obtained with the model and we show the summary of the 
results in Fig. \ref{fig:iipeg_summary}. We sample the temperature maps using again
the HEALPix pixelation with $N_\mathrm{side}=16$. We show results when the 12
rotation phases are used for reconstruction. As well, and to check for the sensitivity
to the number of observations, we also show the results obtained when one out of every two 
observations are used (resulting in only 6 rotation phases) and one out of every 
four (for a total of 3 rotation phases). 
When using the full observed data, a spot very close to the equator is found in all percentiles. 
The median temperature difference between the
spot and the surroundings if $\sim$1000 K. However, there is a large uncertainty in its specific
temperature, between $\sim$3600 K and $\sim$4400 K with 80\% probability. A second less dark
spot is also found close to the northern pole, better visible in phases 0.4-0.6. Its median temperature is
just a few hundred K smaller than the surroundings, although temperatures as low as $\sim$3800 K
cannot be discarded.

The comparison of our results with those found by \cite{rosen15} including all Stokes parameters
shows a similar picture of the star. The polar spot is found in both approaches, although
the Bayesian analysis places the large spot much closer to the equator than \cite{rosen15}.
Anyway, this is compatible with the relatively large uncertainty in the exact location
of the low latitude spot that emerges from the percentiles 10 and 90. We also find
clues of hot spots at intermediate-large latitudes, but their statistical relevance is
small. The same applies to hot spots close to the equator, which are only found
in the percentile 90.

A very similar picture is found when half of the observed rotation phases
are used. This is, in fact, a consequence of the almost equispaced sampling in 
rotation phase, as shown in the left panel of Fig. \ref{fig:iipeg_spectra}. Both dark spots are
placed on exactly the same latitude and longitude. The situation is not so
positive when only three rotation phases are used as observations. The map of the 
median temperature shows a spot that is displaced in longitude with respect to the
previous inferences. However, the percentile 10 indicates the possible presence
of the same equatorial spot found when the full observed material is used. The fact
that it does not show up in the median or percentile 90 maps indicate that this
prediction has a strong bias.

The right panel of Fig. \ref{fig:iipeg_summary} displays samples from the posterior distribution
for the three considered cases. The samples fulfill the summary statistics of Fig. \ref{fig:iipeg_summary}, 
but this helps give an idea of the type of
variability and spatial correlation that we expect to have in the
inference. The equatorial spots appear in almost all the samples when all
available observations are used. The polar spot also appears in many of the
samples. However, the rest of the star is plagued with dark/bright spots
that differ from one sample to the next and behave almost like white noise. 
Only those regions with strong posterior correlations produce structures in the
the percentile maps.

Finally, we carry out a posterior predictive check in Fig. \ref{fig:iipeg_spectra}, where
we show the spectral lines resynthesized in the models from the posterior sampling and 
compare them with the observations. We only show the results for 12 and 6 observations. The
observations are shown in blue. The spectra synthesized in the samples from the posterior
are displayed in orange with transparency, to clearly show that the largest density of
profiles fit the observations. The green line displays the synthetic profiles obtained
from the median estimation of the surface temperature, while the red line shows the median
of all orange profiles. Since radiative transfer is a non-linear process, both profiles
are not the same, though very similar. It is interesting to note that a large majority of the
features are correctly reproduced by the inferred models, although some of the smaller
scale features are not correctly captured by our approximate posterior resulting in a somewhat worse quality of the fit compared to the results presented in \citet{rosen15}.

\section{Conclusions and future work}
We have developed, validated and applied a machine learning model that performs, for the first time, 
fully Bayesian inference of the surface temperature in rotating stars. The model is based on
a normalizing flow that allows us to approximate the posterior distribution in this large-scale
computationally heavy problem. The output of the normalizing flow is conditioned
on a context vector extracted from the observations by a very flexible Transformer
encoder. Likewise, surface temperature maps are compressed with the aid of 
an pre-trained autoencoder.

Although the model is slightly computationally costly to train, this computation
is done only once for a given configuration of observed spectral lines and instrument. Its
amortized character allows us to carry out inference for new observations in fractions of
a second. The resulting posterior distribution for the temperature on the surface
of the star can be exploited to capture uncertainties and correlations. We have used the
median of the marginal distribution per pixel as an estimation of one potential solution, together
with percentiles 10 and 90 as a summary of the uncertainty. Despite these summaries, 
all samples returned by the model are to be seen as potential solutions. Regions with
small variability in the posterior (small difference between the percentiles 10 and 90) constitute
well-constrained values of the surface temperature. Inter-pixel correlations are difficult to
represent in surface data so that we simply show samples from the posterior to give
a limited view of which structures are spatially correlated. We find that our approximation
overestimates the width of the posterior distribution when dealing with spectra
with high $S/N$. We plan to reduce this effect by increasing the size of the 
training set.

There are obvious extensions to this work that we are planning to develop in the
near future. Although requiring minor changes, having a full Bayesian solution
could have a very sizable impact. The first one is the extension to the magnetic case. The
case of the Zeeman DI often requires the use of multi-line techniques because the 
polarimetric signal is well below the noise amplitude for individual lines. The most
widespread multi-line technique is least-squares deconvolution \citep[LSD;][]{donati97,kochukhov10},
which produces a pseudo-line with an increased SNR. Our model can be applied
seamlessly to this case if the training is done with the LSD profiles of synthetic
spectra, in the line with \cite{kochukhov2014}. Instead of one single surface map, 
the Zeeman DI requires the inclusion of the magnetic field vector per pixel in
the star. Since our machine learning model only requires simulations, one can make
the simulated stars as complex as needed or even force some constraints to be fulfilled. For instance, 
one can force the magnetic field to have zero divergence everywhere. If the autoencoder
is properly trained, all solutions will maintain this property. Another possible
extension is to deal with chemical abundance spots such as those observed on early-type stars. Again, given that stars are synthesized
upfront, constraints based on, e.g. atomic diffusion theory, can be imposed, to make them
as realistic as possible.

\begin{acknowledgements}
We acknowledge financial support from the Spanish Ministerio de Ciencia, Innovaci\'on y Universidades through project PGC2018-102108-B-I00 and FEDER funds. 
We also acknowledge support by the Swedish Research Council, the Royal Swedish Academy of Sciences and the Swedish National Space Agency.
%
This project has received funding from the European Research Council (ERC) under the European Union's Horizon 2020 research and innovation program (SUNMAG, grant agreement 759548). The Institute for Solar Physics is supported by a grant for research infrastructures of national importance from the Swedish Research Council (registration number 2017-00625).
This research has made use of NASA's Astrophysics Data System Bibliographic Services.
We acknowledge the community effort devoted to the development of the following 
open-source packages that were
used in this work: \texttt{numpy} \citep[\texttt{numpy.org},][]{numpy20}, 
\texttt{matplotlib} \citep[\texttt{matplotlib.org},][]{matplotlib}, \texttt{PyTorch} 
\citep[\texttt{pytorch.org},][]{pytorch19} and \texttt{zarr} (\texttt{github.com/zarr-developers/zarr-python}).
\end{acknowledgements}


\begin{thebibliography}{55}
    \expandafter\ifx\csname natexlab\endcsname\relax\def\natexlab#1{#1}\fi
    
    \bibitem[{{Adelman} {et~al.}(2002){Adelman}, {Gulliver}, {Kochukhov}, \&
      {Ryabchikova}}]{adelman2002}
    {Adelman}, S.~J., {Gulliver}, A.~F., {Kochukhov}, O.~P., \& {Ryabchikova},
      T.~A. 2002, \apj, 575, 449
    
    \bibitem[{Ba {et~al.}(2016)Ba, Kiros, \& Hinton}]{ba2016layer}
    Ba, J.~L., Kiros, J.~R., \& Hinton, G.~E. 2016 [\eprint[arXiv]{1607.06450}]
    
    \bibitem[{{Bayes}(1764)}]{bayes}
    {Bayes}, T. 1764, Philosophical Transactions of the Royal Society of London,
      53, 370
    
    \bibitem[{Beaumont {et~al.}(2002)Beaumont, Zhang, \& Balding}]{Beaumont02}
    Beaumont, M.~A., Zhang, W., \& Balding, D.~J. 2002, Genetics, 162, 2025
    
    \bibitem[{{Berdyugina}(1998)}]{berdyugina1998}
    {Berdyugina}, S.~V. 1998, \aap, 338, 97
    
    \bibitem[{{Brown} {et~al.}(1991){Brown}, {Donati}, {Rees}, \&
      {Semel}}]{brown91}
    {Brown}, S.~F., {Donati}, J.~F., {Rees}, D.~E., \& {Semel}, M. 1991, \aap, 250,
      463
    
    \bibitem[{{Chan} {et~al.}(2020){Chan}, {Monteiro}, {Kellnhofer}, {Wu}, \&
      {Wetzstein}}]{pigan20}
    {Chan}, E.~R., {Monteiro}, M., {Kellnhofer}, P., {Wu}, J., \& {Wetzstein}, G.
      2020, arXiv e-prints, arXiv:2012.00926
    
    \bibitem[{{Collier Cameron} \& {Horne}(1986)}]{collier-cameron1986}
    {Collier Cameron}, A. \& {Horne}, K.~D. 1986, {Maximum Entropy Reconstruction
      of Starspot Distributions}, ed. M.~{Zeilik} \& D.~M. {Gibson}, Vol. 254, 205
    
    \bibitem[{Cranmer {et~al.}(2020)Cranmer, Brehmer, \& Louppe}]{cranmer20}
    Cranmer, K., Brehmer, J., \& Louppe, G. 2020, Proceedings of the National
      Academy of Sciences, 117, 30055
    
    \bibitem[{{Crossfield} {et~al.}(2014){Crossfield}, {Biller}, {Schlieder},
      {Deacon}, {Bonnefoy}, {Homeier}, {Allard}, {Buenzli}, {Henning}, {Brandner},
      {Goldman}, \& {Kopytova}}]{crossfield2014}
    {Crossfield}, I.~J.~M., {Biller}, B., {Schlieder}, J.~E., {et~al.} 2014, \nat,
      505, 654
    
    \bibitem[{{Deutsch}(1958)}]{deutsch58}
    {Deutsch}, A.~J. 1958, in Electromagnetic Phenomena in Cosmical Physics, ed.
      B.~{Lehnert}, Vol.~6, 209
    
    \bibitem[{{Dinh} {et~al.}(2014){Dinh}, {Krueger}, \& {Bengio}}]{Dinh2014}
    {Dinh}, L., {Krueger}, D., \& {Bengio}, Y. 2014, arXiv e-prints,
      arXiv:1410.8516
    
    \bibitem[{{Donati}(2003)}]{donati03}
    {Donati}, J.~F. 2003, in Astronomical Society of the Pacific Conference Series,
      Vol. 307, Solar Polarization, ed. J.~{Trujillo-Bueno} \& J.~{Sanchez
      Almeida}, 41
    
    \bibitem[{{Donati} {et~al.}(1997){Donati}, {Semel}, {Carter}, {Rees}, \&
      {Collier Cameron}}]{donati97}
    {Donati}, J.~F., {Semel}, M., {Carter}, B.~D., {Rees}, D.~E., \& {Collier
      Cameron}, A. 1997, \mnras, 291, 658
    
    \bibitem[{{Durkan} {et~al.}(2019){Durkan}, {Bekasov}, {Murray}, \&
      {Papamakarios}}]{Durkan2019}
    {Durkan}, C., {Bekasov}, A., {Murray}, I., \& {Papamakarios}, G. 2019, arXiv
      e-prints, arXiv:1906.04032
    
    \bibitem[{Durkan {et~al.}(2020)Durkan, Bekasov, Murray, \&
      Papamakarios}]{nflows}
    Durkan, C., Bekasov, A., Murray, I., \& Papamakarios, G. 2020, {nflows}:
      normalizing flows in {PyTorch}
    
    \bibitem[{{Folsom} {et~al.}(2018){Folsom}, {Bouvier}, {Petit}, {L{\`e}bre},
      {Amard}, {Palacios}, {Morin}, {Donati}, \& {Vidotto}}]{folsom2018}
    {Folsom}, C.~P., {Bouvier}, J., {Petit}, P., {et~al.} 2018, \mnras, 474, 4956
    
    \bibitem[{{Goncharskij} {et~al.}(1982){Goncharskij}, {Stepanov}, {Khokhlova},
      \& {Yagola}}]{goncharsky82}
    {Goncharskij}, A.~V., {Stepanov}, V.~V., {Khokhlova}, V.~L., \& {Yagola}, A.~G.
      1982, \azh, 59, 1146
    
    \bibitem[{{Goncharskij} {et~al.}(1977){Goncharskij}, {Stepanov}, {Kokhlova}, \&
      {Yagola}}]{goncharskii1977}
    {Goncharskij}, A.~V., {Stepanov}, V.~V., {Kokhlova}, V.~L., \& {Yagola}, A.~G.
      1977, Soviet Astronomy Letters, 3, 147
    
    \bibitem[{{Gregory}(2005)}]{gregory05}
    {Gregory}, P.~C. 2005, {Bayesian Logical Data Analysis for the Physical
      Sciences} (Cambridge: Cambridge University Press)
    
    \bibitem[{{Gustafsson} {et~al.}(2008){Gustafsson}, {Edvardsson}, {Eriksson},
      {J{\o}rgensen}, {Nordlund}, \& {Plez}}]{marcs08}
    {Gustafsson}, B., {Edvardsson}, B., {Eriksson}, K., {et~al.} 2008, \aap, 486,
      951
    
    \bibitem[{Harris {et~al.}(2020)Harris, Millman, van~der Walt, Gommers,
      Virtanen, Cournapeau, Wieser, Taylor, Berg, Smith, Kern, Picus, Hoyer, van
      Kerkwijk, Brett, Haldane, Fernández~del Río, Wiebe, Peterson,
      Gérard-Marchant, Sheppard, Reddy, Weckesser, Abbasi, Gohlke, \&
      Oliphant}]{numpy20}
    Harris, C.~R., Millman, K.~J., van~der Walt, S.~J., {et~al.} 2020, Nature, 585,
      357–362
    
    \bibitem[{He {et~al.}(2015)He, Zhang, Ren, \& Sun}]{He2015}
    He, K., Zhang, X., Ren, S., \& Sun, J. 2015, arXiv preprint arXiv:1512.03385
    
    \bibitem[{{Hunter}(2007)}]{matplotlib}
    {Hunter}, J.~D. 2007, Computing in Science Engineering, 9, 90
    
    \bibitem[{{Hussain} {et~al.}(2000){Hussain}, {Donati}, {Collier Cameron}, \&
      {Barnes}}]{hussain2000}
    {Hussain}, G.~A.~J., {Donati}, J.-F., {Collier Cameron}, A., \& {Barnes}, J.~R.
      2000, \mnras, 318, 961
    
    \bibitem[{{Khokhlova}(1976)}]{khokhlova1976}
    {Khokhlova}, V.~L. 1976, \sovast, 19, 576
    
    \bibitem[{{Kingma} \& {Ba}(2014)}]{Kingma2014}
    {Kingma}, D.~P. \& {Ba}, J. 2014, arXiv e-prints, arXiv:1412.6980
    
    \bibitem[{Kingma \& Dhariwal(2018)}]{glow18}
    Kingma, D.~P. \& Dhariwal, P. 2018, in Advances in Neural Information
      Processing Systems 31: Annual Conference on Neural Information Processing
      Systems 2018, 10236--10245
    
    \bibitem[{Kobyzev {et~al.}(2020)Kobyzev, Prince, \& Brubaker}]{flows_kobyzev20}
    Kobyzev, I., Prince, S., \& Brubaker, M. 2020, IEEE Transactions on Pattern
      Analysis and Machine Intelligence, 1
    
    \bibitem[{{Kochukhov}(2004)}]{kochukhov2004}
    {Kochukhov}, O. 2004, \aap, 423, 613
    
    \bibitem[{{Kochukhov}(2016)}]{kochukhov2016}
    {Kochukhov}, O. 2016, in Lecture Notes in Physics, Vol. 914, Lecture Notes in
      Physics, ed. J.-P. {Rozelot} \& C.~{Neiner}, 177--204
    
    \bibitem[{{Kochukhov} {et~al.}(2014){Kochukhov}, {L{\"u}ftinger}, {Neiner},
      {Alecian}, \& {MiMeS Collaboration}}]{kochukhov2014}
    {Kochukhov}, O., {L{\"u}ftinger}, T., {Neiner}, C., {Alecian}, E., \& {MiMeS
      Collaboration}. 2014, \aap, 565, A83
    
    \bibitem[{{Kochukhov} {et~al.}(2010){Kochukhov}, {Makaganiuk}, \&
      {Piskunov}}]{kochukhov10}
    {Kochukhov}, O., {Makaganiuk}, V., \& {Piskunov}, N. 2010, \aap, 524, A5
    
    \bibitem[{{Kochukhov} {et~al.}(2019){Kochukhov}, {Shultz}, \&
      {Neiner}}]{kochukhov2019}
    {Kochukhov}, O., {Shultz}, M., \& {Neiner}, C. 2019, \aap, 621, A47
    
    \bibitem[{{Krachmalnicoff} \& {Tomasi}(2019)}]{2019A&A...628A.129K}
    {Krachmalnicoff}, N. \& {Tomasi}, M. 2019, \aap, 628, A129
    
    \bibitem[{Nair \& Hinton(2010)}]{relu10}
    Nair, V. \& Hinton, G.~E. 2010, in Proceedings of the 27th International
      Conference on Machine Learning (ICML-10), June 21-24, 2010, Haifa, Israel,
      807--814
    
    \bibitem[{{Paszke} {et~al.}(2019){Paszke}, {Gross}, {Massa}, {Lerer},
      {Bradbury}, {Chanan}, {Killeen}, {Lin}, {Gimelshein}, {Antiga}, {Desmaison},
      {K{\"o}pf}, {Yang}, {DeVito}, {Raison}, {Tejani}, {Chilamkurthy}, {Steiner},
      {Fang}, {Bai}, \& {Chintala}}]{PyTorch}
    {Paszke}, A., {Gross}, S., {Massa}, F., {et~al.} 2019, arXiv e-prints,
      arXiv:1912.01703
    
    \bibitem[{Paszke {et~al.}(2019)Paszke, Gross, Massa, Lerer, Bradbury, Chanan,
      Killeen, Lin, Gimelshein, Antiga, Desmaison, Kopf, Yang, DeVito, Raison,
      Tejani, Chilamkurthy, Steiner, Fang, Bai, \& Chintala}]{pytorch19}
    Paszke, A., Gross, S., Massa, F., {et~al.} 2019, in Advances in Neural
      Information Processing Systems 32, ed. H.~Wallach, H.~Larochelle,
      A.~Beygelzimer, F.~d'\'{e} Buc, E.~Fox, \& R.~Garnett (Curran Associates,
      Inc.), 8024--8035
    
    \bibitem[{{Pearson}(1901)}]{pearson01}
    {Pearson}, K. 1901, Philosophical Magazine, 2, 559
    
    \bibitem[{{Perez} {et~al.}(2017){Perez}, {Strub}, {de Vries}, {Dumoulin}, \&
      {Courville}}]{film17}
    {Perez}, E., {Strub}, F., {de Vries}, H., {Dumoulin}, V., \& {Courville}, A.
      2017, arXiv e-prints, arXiv:1709.07871
    
    \bibitem[{{Piskunov} \& {Kochukhov}(2002)}]{piskunov2002}
    {Piskunov}, N. \& {Kochukhov}, O. 2002, \aap, 381, 736
    
    \bibitem[{{Piskunov} \& {Khokhlova}(1983)}]{piskunov1983}
    {Piskunov}, N.~E. \& {Khokhlova}, V.~L. 1983, Soviet Astronomy Letters, 9, 346
    
    \bibitem[{{Piskunov} {et~al.}(1990){Piskunov}, {Tuominen}, \&
      {Vilhu}}]{piskunov1990}
    {Piskunov}, N.~E., {Tuominen}, I., \& {Vilhu}, O. 1990, \aap, 230, 363
    
    \bibitem[{{Planck Collaboration} {et~al.}(2020){Planck Collaboration},
      {Aghanim}, {Akrami}, {Arroja}, {Ashdown}, {Aumont}, {Baccigalupi},
      {Ballardini}, {Banday}, {Barreiro}, {Bartolo}, {Basak}, {Battye}, {Benabed},
      {Bernard}, {Bersanelli}, {Bielewicz}, {Bock}, {Bond}, {Borrill}, {Bouchet},
      {Boulanger}, {Bucher}, {Burigana}, {Butler}, {Calabrese}, {Cardoso},
      {Carron}, {Casaponsa}, {Challinor}, {Chiang}, {Colombo}, {Combet},
      {Contreras}, {Crill}, {Cuttaia}, {de Bernardis}, {de Zotti}, {Delabrouille},
      {Delouis}, {D{\'e}sert}, {Di Valentino}, {Dickinson}, {Diego}, {Donzelli},
      {Dor{\'e}}, {Douspis}, {Ducout}, {Dupac}, {Efstathiou}, {Elsner},
      {En{\ss}lin}, {Eriksen}, {Falgarone}, {Fantaye}, {Fergusson},
      {Fernandez-Cobos}, {Finelli}, {Forastieri}, {Frailis}, {Franceschi},
      {Frolov}, {Galeotta}, {Galli}, {Ganga}, {G{\'e}nova-Santos}, {Gerbino},
      {Ghosh}, {Gonz{\'a}lez-Nuevo}, {G{\'o}rski}, {Gratton}, {Gruppuso},
      {Gudmundsson}, {Hamann}, {Handley}, {Hansen}, {Helou}, {Herranz},
      {Hildebrandt}, {Hivon}, {Huang}, {Jaffe}, {Jones}, {Karakci}, {Keih{\"a}nen},
      {Keskitalo}, {Kiiveri}, {Kim}, {Kisner}, {Knox}, {Krachmalnicoff}, {Kunz},
      {Kurki-Suonio}, {Lagache}, {Lamarre}, {Langer}, {Lasenby}, {Lattanzi},
      {Lawrence}, {Le Jeune}, {Leahy}, {Lesgourgues}, {Levrier}, {Lewis},
      {Liguori}, {Lilje}, {Lilley}, {Lindholm}, {L{\'o}pez-Caniego}, {Lubin}, {Ma},
      {Mac{\'\i}as-P{\'e}rez}, {Maggio}, {Maino}, {Mandolesi}, {Mangilli},
      {Marcos-Caballero}, {Maris}, {Martin}, {Martinelli},
      {Mart{\'\i}nez-Gonz{\'a}lez}, {Matarrese}, {Mauri}, {McEwen}, {Meerburg},
      {Meinhold}, {Melchiorri}, {Mennella}, {Migliaccio}, {Millea}, {Mitra},
      {Miville-Desch{\^e}nes}, {Molinari}, {Moneti}, {Montier}, {Morgante}, {Moss},
      {Mottet}, {M{\"u}nchmeyer}, {Natoli}, {N{\o}rgaard-Nielsen}, {Oxborrow},
      {Pagano}, {Paoletti}, {Partridge}, {Patanchon}, {Pearson}, {Peel}, {Peiris},
      {Perrotta}, {Pettorino}, {Piacentini}, {Polastri}, {Polenta}, {Puget},
      {Rachen}, {Reinecke}, {Remazeilles}, {Renault}, {Renzi}, {Rocha}, {Rosset},
      {Roudier}, {Rubi{\~n}o-Mart{\'\i}n}, {Ruiz-Granados}, {Salvati}, {Sandri},
      {Savelainen}, {Scott}, {Shellard}, {Shiraishi}, {Sirignano}, {Sirri},
      {Spencer}, {Sunyaev}, {Suur-Uski}, {Tauber}, {Tavagnacco}, {Tenti},
      {Terenzi}, {Toffolatti}, {Tomasi}, {Trombetti}, {Valiviita}, {Van Tent},
      {Vibert}, {Vielva}, {Villa}, {Vittorio}, {Wandelt}, {Wehus}, {White},
      {White}, {Zacchei}, \& {Zonca}}]{2020A&A...641A...1P}
    {Planck Collaboration}, {Aghanim}, N., {Akrami}, Y., {et~al.} 2020, \aap, 641,
      A1
    
    \bibitem[{{Rice}(2002)}]{rice2002}
    {Rice}, J.~B. 2002, Astronomische Nachrichten, 323, 220
    
    \bibitem[{{Rice} \& {Strassmeier}(2000)}]{rice2000}
    {Rice}, J.~B. \& {Strassmeier}, K.~G. 2000, \aaps, 147, 151
    
    \bibitem[{{Ros{\'e}n} {et~al.}(2015){Ros{\'e}n}, {Kochukhov}, \&
      {Wade}}]{rosen15}
    {Ros{\'e}n}, L., {Kochukhov}, O., \& {Wade}, G.~A. 2015, \apj, 805, 169
    
    \bibitem[{Rubin(1984)}]{rubin84}
    Rubin, D.~B. 1984, The Annals of Statistics, 12, 1151
    
    \bibitem[{{Semel}(1989)}]{semel89}
    {Semel}, M. 1989, \aap, 225, 456
    
    \bibitem[{Sitzmann {et~al.}(2020)Sitzmann, Martel, Bergman, Lindell, \&
      Wetzstein}]{sitzmann2020implicit}
    Sitzmann, V., Martel, J.~N., Bergman, A.~W., Lindell, D.~B., \& Wetzstein, G.
      2020, arXiv preprint arXiv:2006.09661
    
    \bibitem[{{Skilling} \& {Bryan}(1984)}]{skilling84}
    {Skilling}, J. \& {Bryan}, R.~K. 1984, \mnras, 211, 111
    
    \bibitem[{{Strassmeier} {et~al.}(2019){Strassmeier}, {Carroll}, \&
      {Ilyin}}]{strassmeier2019}
    {Strassmeier}, K.~G., {Carroll}, T.~A., \& {Ilyin}, I.~V. 2019, \aap, 625, A27
    
    \bibitem[{{Tikhonov} \& {Arsenin}(1977)}]{tikhonov1977}
    {Tikhonov}, A.~N. \& {Arsenin}, V.~Y. 1977, {Solution of Ill-posed Problems}
      (Wiley: New York)
    
    \bibitem[{Vaswani {et~al.}(2017)Vaswani, Shazeer, Parmar, Uszkoreit, Jones,
      Gomez, Kaiser, \& Polosukhin}]{Transformer17}
    Vaswani, A., Shazeer, N., Parmar, N., {et~al.} 2017, in Proceedings of the 31st
      International Conference on Neural Information Processing Systems, NIPS'17,
      6000–6010
    
    \bibitem[{{Vogt} {et~al.}(1987){Vogt}, {Penrod}, \& {Hatzes}}]{vogt87}
    {Vogt}, S.~S., {Penrod}, G.~D., \& {Hatzes}, A.~P. 1987, \apj, 321, 496
    
    \end{thebibliography}
\end{document}